\documentstyle[preprint,tighten,aps,epsfig]{revtex}
\def\qed{\hbox{${\vcenter{\vbox{                          
   \hrule height 0.4pt\hbox{\vrule width 0.4pt height 6pt
   \kern5pt\vrule width 0.4pt}\hrule height 0.4pt}}}$}}
\newcommand{\bea}{\begin{eqnarray}}
\newcommand{\eea}{\end{eqnarray}}
\newcommand{\be}{\begin{equation}}
\newcommand{\ee}{\end{equation}}

\def\shiftleft#1{#1\llap{#1\hskip 0.04em}}
\def\shiftdown#1{#1\llap{\lower.04ex\hbox{#1}}}
\def\thick#1{\shiftdown{\shiftleft{#1}}}
\def\b#1{\thick{\hbox{$#1$}}}
\newcommand{\titj}{{\mbox{\boldmath $\tau$}}_i\cdot{\mbox{\boldmath$\tau$}}_j}
\newcommand{\sisj}{{\mbox{\boldmath $\sigma$}}_i\cdot{\mbox{\boldmath $
            \sigma$}}_j}
\newcommand{\lilj}{\lambda_i^a\cdot\lambda_j^a}

\begin{document}
\draft

\preprint{\vbox{\hfill ADP-98-5/T285  }}
\title{The $d'$-Dibaryon in a Colored Cluster Model}  
\author{A. J.\ Buchmann$^1,$\footnote{Electronic address:
        alfons.buchmann@uni-tuebingen.de}, 
        Georg Wagner$^{1,2,}$\footnote{Electronic address: 
        gwagner@physics.adelaide.edu.au}
        and Amand Faessler$^1,$\footnote{Electronic address:
        amand.faessler@uni-tuebingen.de} }
\address{$^1$Institut f\"{u}r Theoretische Physik, Universit\"at T\"ubingen,
         Auf der Morgenstelle 14, D-72076 T\"ubingen, Germany }
\address{$^2$Centre for the Subatomic Structure of Matter (CSSM),
         University of Adelaide, Australia 5005. }


\maketitle

\begin{abstract} 
The mass and wave function of a six-quark system with
quantum numbers J$^P$=0$^-$, T=0, called $d'$, are calculated. 
We use a colored diquark-tetraquark cluster model for the six-quark wave 
function. A constituent quark model Hamiltonian with a
two-body confinement potential, and residual one-gluon, one-pion, 
and one-sigma exchange interactions is used. The complications due to 
the quark exchange interactions between tetraquark and diquark 
clusters (Pauli principle) are taken into account within the framework of the 
Resonating Group Method. The calculated $d'$ mass is some 350 MeV above 
the empirical value if the same two-body confinement strength 
as in the  nucleon and $\Delta$ is used.
This paper also examines the validity of the usual assumption of a  
universal two-quark confinement strength. We propose that 
the effective two-body confinement strength in an exotic six-quark system,
such as the $d'$, could be weaker than in a single baryon. 
The weaker confinement hypothesis leads to a $d'$ mass of $M_{d'}=2092$ MeV
and a $d'$ radius of $r_{d'}=1.53$ fm.

\end{abstract}

\smallskip
\pacs{PACS number(s): 14.20.Pt, 12.39.Pn, 12.39.Jh, 12.40.Yx} 

\section{Introduction}
\label{sec:intro} 
  
First indications for the existence of the $d'$ dibaryon 
came from the narrow peak observed in the pionic ($\pi^+$,$\pi^-$) double
charge exchange (DCX) reaction on nuclei.
At an incident pion energy 
of $T_{\pi}\sim$ 50 MeV and forward pion scattering angles $\Theta=5^o$,
the DCX cross section, for example for 
$\pi^+\; +\;  ^{12}\!C \; \to \;   ^{12}\!O \; +\;  \pi^{-}$,
displays a narrow peak \cite{bil92,mar91,bil93}.
Because of charge conservation, the DCX reaction involves at least 
two nucleons within the nucleus, and it has been shown that the DCX reaction
is therefore very sensitive to 
short-range NN-correlations \cite{ble87}.                
The possible importance of explicit quark degrees of freedom  
in DCX at 50 MeV and forward angles 
was pointed out some time ago \cite{mil87}.           
In the meantime, dedicated DCX experiments \cite{bil93,lei89,foe97} 
on a large number of light and medium heavy nuclei 
ranging from $^7$Li to $^{56}$Fe
 have unambiguously confirmed  
the existence of a narrow resonance-like structure
at this pion energy and angle.

While conventional DCX calculations  \cite{aue88,kag94} 
have so far been unable to explain these experimental results,
the assumption of a single narrow 
baryon number $B=2$ resonance with total spin, parity J$^P$=0$^-$, 
isospin T=even, a resonance energy of $M_{d'}$= 2065 MeV, and 
a free decay width of 
$\Gamma_{\pi NN} \approx$ 0.5 MeV works
extremely well in describing all available DCX data \cite{bil93,wag94}.
To exclude a possible nuclear structure explanation \cite{sim95}, 
experimentalists have searched for the $d'$ in proton-proton collisions, e.g. 
in  $p\; p \; \longrightarrow\; p\; p \; \pi^+ \; \pi^-.$  
A 4$\sigma$ enhancement over the background exactly at the $d'$ 
position has been observed in the $\pi^-pp$ invariant mass spectrum 
\cite{bro96}. 

Two alternative models for the structure of the $d'$ have been 
discussed in the literature. If the $d'$ is predominantly composed of
two {\it colorless} three-quark clusters, it could be 
a resonance in the $NN^*$ channel, where $N^*$ is the first 
negative parity excitation of the nucleon at 1535 MeV. 
Because the empirical mass of the $d'$ is far below 
the $N^*N$ threshold, this requires an enormous 
binding energy for the bound $NN^*$ system. This interpretation is reminiscent 
of the deep-lying bound states in the Moscow NN potential \cite{kuk98}.
The $d'$-dibaryon has also been interpreted as an isospin 
$T=2$ $N\Delta$ resonance \cite{val95}, 
a $\pi NN$ state with $T=2$ \cite{gar94}, and most recently as a  
 $\pi$NN state in the isospin $T=0$ channel\cite{gar97}, 
located at 2018 MeV.

It is important to note that due to its quantum numbers 
($L=1$, $S=1$, $T=0$), the $d'$ cannot be composed of two ground state 
nucleons, because such a state is Pauli-forbidden. The Pauli principle 
demands that the quantum numbers of a two-nucleon system 
satisfy $(-1)^{L+S+T}=-1$.  Therefore, the $d'$ cannot simply decay into 
two nucleons. A decay into two nucleons is possible only if a pion or a 
photon is emitted simultaneously. The correspondingly small phase space 
naturally explains the small width of the $d'$.

On the other hand, quantum chromodynamics (QCD) does not preclude
the possibility that the quarks in the initial $\pi NN$ system 
rearrange themselves into energetically favorable but unobservable 
{\it colored} clusters \cite{set85}. This idea
is not new. Several scenarios for the arrangement of quarks and antiquarks 
into colored clusters combining to an overall color neutral $B=2$ 
system have been proposed \cite{jaf77,sil92,lom90,Ans93}. An early 
suggestion is the `demon' deuteron, 
which is a $J^P=0^-$, $T=0$ dibaryon consisting of three pairs of diquarks 
\cite{Fre82}. 
Other partitionings have been studied in   
the stringlike bag model \cite{mul80,kon87}.  This model predicts that 
a rotating ($L=1$) dumbbell-like configuration with a colored diquark ($q^2$)
and tetraquark ($q^4$) cluster connected by a flux tube of color-electric 
field lines is the state with lowest energy 
for a six-quark system with $d'$ quantum numbers.  
Fig.\ \ref{fig1} shows how such 
a $q^2-q^4$ cluster system could be formed from the initial $\pi$NN state.

A shortcoming of the stringlike bag model \cite{mul80,kon87} 
is that it employs rigid impenetrable tetraquark and diquark clusters
at the ends of a rotating color string. This model neither allows that the 
clusters merge into a single compound six-quark bag nor that they exchange 
quarks. Only the quarks within the individual clusters are 
antisymmetrized but not the quarks belonging to different clusters. 
Therefore, the stringlike bag model does not fully respect the Pauli 
principle. The stringlike bag model is a good approximation only 
for high angular momentum states ($L > 5$) \cite{joh76}. In this 
case, the system is fairly elongated because of centrifugal forces, and the  
probability of cluster overlap is small. However, for a low lying $L=1$ 
excitation, such as the $d'$, one expects considerable overlap
between the clusters and a substantial amount of quark 
exchange between them. We are sceptical that an $L=1$ state is sufficiently  
stretched for the stringlike bag model treatment to be valid.
The condition of validity is $l > 2R_0$ \cite{mul80}, where $l$ 
is the length and $R_0$ the radius of the color flux tube connecting the 
colored quark clusters. For the $d'$, this condition is barely 
satisfied, and one anticipates corrections to the bag model
prediction $M_{d'} \approx 2100$ MeV \cite{mul80,kon87}.

The purpose of this work is to study 
the dynamics of diquark-tetraquark 
relative motion using a microscopic quark model Hamiltonian and 
the Resonating Group Method (RGM). The model used here 
accurately reproduces the mass 
of the deuteron, which is the only established dibaryon. 
By comparing the RGM solutions with our previous results 
\cite{glo94,wag95,ito96}, employing a single $s^5p^1$ 
six-quark ``bag'', we can test the validity of the assumption
underlying the stringlike bag model, namely
that the $d'$ is a stretched $q^2-q^4$ system.
In the present work, 
all complications arising from the quark exchange 
interactions (Pauli principle) are rigorously taken into account,
and their effect on the $d'$ mass and wave function is investigated in 
some detail.
In addition,  the present paper critically examines
the validity of the assumption of a universal two-body confinement
strength and explores how a possible 
reduction of the effective confinement strength 
in a compound six-quark system
affects the mass and the size of the $d'$. 

The paper is organized as follows.
Section \ref{sec:nrqm} provides a short description of the chiral quark model 
used. Section \ref{sec:rgm} presents the six-quark Resonating Group Method
approach.
Section \ref{sec:res} discusses the numerical results for the mass
and size of the $d'$ for several confinement models and compares
the present predictions to those of the single-bag shell model 
and the stringlike bag model. 
Our summary and conclusions are given in Section \ref{sec:sum}.
An appendix contains explicit  expressions of the 
norm  and Hamiltonian kernels needed for the 
solution of the RGM equation of motion. 

\section{The chiral constituent quark model} 
\label{sec:nrqm} 

In the chiral constituent quark model 
a system of $n$-quarks with equal masses $m_q=313$ MeV=$m_N$/3 
(SU$_F$(2)) is described by the Hamiltonian  
\begin{equation}
\label{Ham}
  H = \sum_{i=1}^{n} \Bigl ( m_q+ {{\bf p}^{2}_{i}\over 2m_{q}} \Bigr )
  -{{\bf P}^2\over n(2m_q)}
  + \sum_{i<j}^{n} V^{\rm{conf}}({\bf r}_i,{\bf r}_j) +
  \sum_{i<j}^{n} V^{\rm{res}}({\bf r}_i,{\bf r}_j),  
\end{equation}
where ${\bf r}_i, {\bf p}_i$
are the spatial and momentum coordinates of the i-th
quark and ${\bf P}$ is the total momentum of the
$n$-quark system. The exact removal of the center of mass kinetic energy 
by the third term represents an important advantage of the present approach.

The residual interactions 
$V^{\rm{res}}=V^{\rm{OGEP}}+V^{\rm{OPEP}}+V^{\rm{OSEP}}$ 
between quarks shown in Fig.\ \ref{fig2}
model the most relevant properties of QCD, such as asymptotic freedom at 
high, and spontaneous chiral symmetry breaking at low energies. 
The one-gluon exchange potential \cite{ruj75} 
\begin{equation}
\label{gluon}
  V^{\rm{OGEP}} ({\bf r}_i,{\bf r}_j)  =  {\alpha_{s}\over 4}
  \b{\lambda}_i\cdot\b{\lambda}_j \Biggl \lbrace
  {1\over r}- {\pi\over m_q^2} \left ( 1+{2\over 3}
  \b{\sigma}_{i}\cdot\b{\sigma}_{j} \right ) \delta({\bf r}) \Biggr \rbrace
\end{equation}
provides an effective quark-quark interaction that
has the spin-color structure of QCD at short distances.
Here, ${\bf r}={\bf r}_i-{\bf r}_j$, and $\b{\sigma}_i$ is the usual 
Pauli spin matrix. The $\b{\lambda}_i$ represent the Gell-Mann matrices 
of SU(3)$_{\rm{color}}$.
We neglect tensor and spin-orbit forces, 
which have been shown to be of minor importance
for the $d'$ \cite{wag95}. Only central potentials are considered
in this work.

The spontaneous breaking of chiral symmetry of low-energy QCD 
by the physical vacuum is responsible for the constituent 
quark mass generation \cite{man84}, as well as for the appearence of 
pseudoscalar and scalar collective excitations of the vacuum 
($\pi$ and $\sigma$ fields).
These collective degrees of freedom couple 
directly to the constituent quarks. 
In the present quark model with two quark flavors, 
this mechanism is modelled by regularized one-pion and one-sigma exchange 
potentials between constituent quarks \cite{obu90,fer93,val94}:
\begin{eqnarray}
  V^{\rm{OPEP}}({\bf r}_i,{\bf r}_j) & = &
  {g_{\pi q}^2\over 4 \pi }  {1\over  {12 m_q^2} }
  {\Lambda^2_\pi\over {\Lambda^2_\pi-m_{\pi}^2}} 
  \b{\tau}_{i}\cdot \b{\tau}_{j}\, 
  \b{\sigma}_{i} \cdot \b{\sigma}_{j} \b{\nabla}_{r}^2
  \left ( {e^{-m_{\pi} r}\over r}- {e^{-\Lambda_\pi r}\over r} \right ), 
\label{Pion} \\
  V^{\rm{OSEP}}({\bf r}_i,{\bf r}_j) &= & -{g_{\sigma q}^2\over {4 \pi}} 
  {\Lambda^2_\sigma \over {\Lambda^2_\sigma-m_{\sigma}^2}} 
  \left ( {e^{-m_{\sigma} r}\over r}- {e^{-\Lambda_\sigma r}\over r} \right ),
\label{Sigma}  
\end{eqnarray}
with
\begin{equation}
  {g^2_{\sigma q}\over {4 \pi}} = {g^2_{\pi q}\over {4 \pi}}, \qquad 
  m_{\sigma}^2 \approx (2 m_q)^2+ m_{\pi}^2,  \qquad
  \Lambda_{\pi} =  {\Lambda_{\sigma}}\equiv \Lambda.  
\label{chiral}
\end{equation}
The $\pi q$ coupling constant $g^2_{\pi q}/4 \pi$  is 
related to the $\pi N$ coupling constant $g^2_{\pi N}/4 \pi$ 
by $g_{\pi q} = (3/5)(m_q/m_N) g_{\pi N}$ with $g_{\pi N}=13.19$ \cite{ber90}.
The $\pi q$ cut-off mass $\Lambda$ describes the finite size of the 
constituent quark due to its pion cloud
\be
\label{cqs}
r_q^2= {3 \over \Lambda^2}. 
\ee
Here, we choose $\Lambda=4.2$ fm$^{-1}$ which gives $r_q=0.41$ fm,
and leads to a soft $\pi N$ form factor \cite{buc91}. The sigma parameters 
are fixed by Eq.(\ref{chiral}). In particular we use $m_{\sigma}=626$ MeV. 

The usefulness and 
phenomenological success of the model has previously been demonstrated. 
For example, the chiral constituent quark model has 
recently been used to calculate the positive 
parity spectrum of the nucleon and various electromagnetic
observables of the ground state octet and decuplet baryons and 
their excited states \cite{buc91}. It has also
been applied to calculate the properties of the deuteron \cite{val94},
NN phase shifts \cite{fer93,val94}, hyperon-hyperon interactions,
as well as the H-particle \cite{str90}.

\subsection{Confinement models} 
\label{subsec:conf} 

There are mainly two types of confinement potentials discussed 
in the literature, namely the two-body confinement potential 
introduced by Lipkin \cite{Lip73}, and the color flux tube (string)
model of confinement. In the color flux tube model, the color string 
describes the hidden gluon degrees of freedom that are necessary to preserve 
color gauge invariance of the underlying field theory. In a many quark system, 
or in the interaction region of hadron-hadron collisions, the strings can 
change their positions and oscillate between different configurations. 
This continuous flipping of color strings can be effectively described 
in the flip--flop model \cite{Len86}, which has been introduced to avoid 
the long-range color van der Waals forces in hadron-hadron interactions.
In the flip-flop model, the confinement interaction between any pair of 
quarks depends on the position of the remaining quarks; it thus contains
many-body operators.

Here, for reasons of simplicity, we consider several 
{\it two-body} confinement potential 
models of Lipkin-type, which differ in their radial 
dependence (see Fig.3) but not in their color structure. 
We should keep in mind that these 
potentials may not be adequate for describing 
the complicated dynamics of changing
color strings in a many-quark system.

In the constituent quark model one often takes a simple
quadratic confinement potential:
\begin{equation}
  V^{\rm{conf}}_q({\bf r}_i,{\bf r}_j)   = 
          -a_c^{(q)} \b{\lambda}_i\cdot \b{\lambda}_j
           ({\bf r }_i-{\bf r }_j)^2 \; . 
\label{quaconf} 
\end{equation}
Later we use this model to 
argue that the usual assumption of a 
universal two-body confinement strength $a_c^{(q)}$  is too restrictive
and may not be adequate for compact six-quark systems.

Lattice-QCD calculations find that the quark-antiquark 
potential of QCD is linear:
\label{confabcd}
\begin{equation}
  V^{\rm{conf}}_l({\bf r}_i,{\bf r}_j)   = 
          -a_c^{(l)} \b{\lambda}_i\cdot \b{\lambda}_j \,
           \vert ({\bf r }_i-{\bf r }_j) \vert \; , 
\label{linconf} 
\end{equation}
The linear form has also been used in our previous calculation 
\cite{glo94,wag95} using the Translationally Invariant Shell Model 
(TISM).

Another interesting parametrization is the 
$r^{2/3}$ confinement potential:
\begin{equation}
  V^{\rm{conf}}_{\rm{r}}({\bf r}_i,{\bf r}_j)   = 
          -a_c^{({\rm{r}})} \b{\lambda}_i\cdot \b{\lambda}_j
           \vert ({\bf r }_i-{\bf r }_j) \vert^{2/3} \quad .
\label{regconf} 
\end{equation}
In Ref. \cite{fab88} it is shown that 
a $r^{2/3}$ confinement potential and a color-Coulomb potential 
leads in the framework of the nonrelativistic Schr\"{o}dinger equation to   
the observed linear Regge trajectories of hadron masses.

Finally, we consider the color-screened  error-function confinement
\cite{lae86}:
\begin{equation}
  V^{\rm{conf}}_{\rm{e}} ({\bf r}_i,{\bf r}_j)   = 
          -a_c^{({\rm{e}})} \b{\lambda}_i\cdot \b{\lambda}_j
          {\rm{erf}}(\mu r), \quad  \quad
          {\rm{erf}}(x) = \frac{2}{\sqrt{\pi}} \int_0^x
                              e^{-z^2} dz \; .
\label{erfconf} 
\end{equation}
This potential rises linearly for small $r$, but 
as a result of quark-antiquark pair creation  (color screening) grows 
only weakly for intermediate $r$ and finally goes to a constant value at 
large $r$ (see Fig. 3). Lattice calculations show \cite{lae86}  
that such a  behavior of the effective quark-quark potential 
arises if quark-antiquark loops are taken into account.
The inverse of $\mu$ is called color-screening length for which we take
$1/\mu=0.8$ fm. This potential has recently been used by Zhang {\it et al.}  
\cite{zha93}. Alternatively, an exponential form 
$V^{conf}_{exp}=(1 -\exp(-\mu r^2))$  
that matches the $r^2$ behavior for small $r$ has been suggested 
and used by Wang {\it et al.} \cite{Wan92}.

\subsection{Determination of the parameters} 
\label{subsec:para} 

\par
For the nucleon and $\Delta$ wave functions, we use $s^3$ harmonic oscillator 
ground state wave functions $\Phi_{N(\Delta)}$ 
\be
\label{ostcwf}
\mid \Phi_{N (\Delta)} >=
\left ({1 \over \sqrt{3} \pi b_3^2} \right )^{3/2}
\exp \left (- \left (\b{\rho}^2/(4b_3^2)+\b{\lambda}^2/(3b_3^2) 
\right ) \right )
\mid ST>^{N(\Delta)} \times
\mid [111]>^{N(\Delta)}_{color},
\ee
where the Jacobi coordinates $\b{\rho}$ and $\b{\lambda}$ are defined
as $\b{\rho}={\bf r}_1-{\bf r}_2$ and $\b{\lambda}={\bf r}_3-({\bf
r}_1+{\bf r}_2)/2$.
The harmonic oscillator parameter $b_3$ describes the mean square (ms) matter 
radius of the nucleon. The matter radius is defined with respect to
the center of mass coordinate ${\bf R}_N$ of the nucleon and is given as
$$ r_N^2 = \left \langle \Phi_{N}  \left \vert   {1 \over 3} 
\sum_{i=1}^{3} ({\bf r}_i-{\bf R}_N)^2 \right \vert  
\Phi_{N } \right \rangle =  b_3^2. 
$$
As usual, the parameters of the Hamiltonian of Eq.\ (\ref{Ham}) are fitted to
the nucleon and $\Delta$ ground state masses. 
The masses of the N and $\Delta$ are   
calculated as expectation values of the Hamiltonian of Eq.\ (\ref{Ham}) 
between baryon ground state wave functions $\Phi_{N(\Delta)}$.              
Imposing $m_q$=313 MeV and $\Lambda$=4.2 fm$^{-1}$, the parameters 
$a_c^{(3)}$, $\alpha_s$, and $b_3$ are determined by the requirement that
the $N$(939) and $\Delta$(1232) masses are reproduced,
and that the nucleon mass is stable with respect to variations in $b_3$:
\begin{equation}
\label{constraints}
  M_N(b_3)=3m_q = {\rm{939 \,MeV}}, \quad
  M_{\Delta}-M_N = {\rm{293\, MeV}}, \quad
  {\partial M_N(b_3) \over \partial b_3} = 0. 
\end{equation}

Although the three coupled equations (\ref{constraints}) 
for the parameters $a_c^{(3)}$, $\alpha_s$, and  $b_3$  
are non-linear, $a_c^{(3)}$ is predominantly determined by the 
requirement that the nucleon has its physical mass $M_N=939$ MeV,
$\alpha_s$ by the mass difference $M_{\Delta}-M_N$, and 
$b_3$ by the stability condition $\partial M_N(b_3) / \partial b_3 =0$.
The parameters for the different confinement models as determined by 
Eq.\ (\ref{constraints}) are shown in Table \ref{table:para}. 
Parameter sets I-III are obtained with the quadratic confinement
of Eq.\ (\ref{quaconf}) and sets IV,V,VI are obtained  with the linear,
$r^{2/3}$, and error-function confinement of Eqs.(\ref{linconf}-\ref{erfconf})
respectively. 

One can qualitatively understand the trend in the values of the 
parameters $a_c^{(3)}$, $\alpha_s$, and $b_3$ for the different 
confinement models by considering the sequence quadratic (set II), 
linear (set IV), $r^{2/3}$ (set V) and error-function (set VI) confinement 
for the case of the nucleon (see  Fig. \ref{fig3}). One observes that the 
stronger the confinement potential grows at small distances, the further out  
the nucleon wave function (larger $b_3$) extends, 
in order to  minimize the nucleon mass.
With increasing $b_3$, the importance of the one-pion-exchange interaction 
decreases, and the contribution of the one-gluon-exchange 
to the $N\Delta$ mass splitting, --hence $\alpha_s$ -- 
must increase correspondingly.
Finally, because the kinetic energy contribution to the
nucleon mass is lower for larger nucleon sizes ($b_3$), 
a bigger confinement strength $a_c^{(3)}$ is needed to fit the
experimental nucleon mass $M_N$=939 MeV. 

\section{A colored diquark-tetraquark cluster model of the $d'$ dibaryon} 
\label{sec:rgm}

\par

In the DCX reaction, the incident pion and the 
two correlated nucleons may come sufficiently close for the quarks to make
use of the additional possibilities provided by their 
color degree of freedom. In the interaction region, there are 
several possibilities to make an overall color-singlet. The quarks could 
clusterize into two color singlets, two color triplets,  two color sextets, 
and two color octets:

$$
    ({\bf 1} \otimes {\bf 1})^1; \ \ \
    ({\bf {\bar 3}} \otimes {\bf 3})^1; \ \ \
    ({\bf 6} \otimes {\bf {\bar 6}})^1; \ \ \
    ({\bf 8} \otimes {\bf 8})^1.
$$

We recall that a  $q^3-q^3$ clusterization of the $d'$ 
with {\it colorless} three-quark 
clusters is either not allowed by quantum numbers (Pauli principle) or by 
energy considerations. For example, a $N(939)N(939)$ clusterization with
relative angular momentum $L=1$ is not allowed by the Pauli principle.
Similarly, an $N\Delta$ system with $L=1$ 
can only have isospin $T=2$, while present experimental results
seem to favor the $T=0$ assignement.
Furthermore, an $L=0$ $N(939)N^*(1535)$ system, 
where the negative parity 
resides inside one cluster, is far above the experimental  $d'$ mass.
A recent quark cluster model calculation has found no evidence 
for a deeply bound state in this system \cite{Gro96}. 
Therefore, a $N(939)N^*(1535)$ dibaryon decays strongly into $NN\pi$. 
It has a large width and thus cannot be a viable candidate for the $d'$.
Thus, the $d'$ is presumably not a state which 
is predominantly composed of two colorless three-quark clusters. 
 
In the stringlike bag model, the  
color-triplet (${\bf {3}}$) tetraquark, 
color-antitriplet (${\bf {\bar 3}}$) diquark 
clusterization is the energetically most 
favorable configuration for a system 
with $d'$ quantum numbers. For such a system, a relatively small mass of 
$M_{d'} \approx 2100 $ MeV has been obtained \cite{mul80}.

The six-quark wave function for the $d'$-dibaryon 
is expanded into the ${\bf 3} \otimes {\bf {\bar 3}} $ 
tetraquark-diquark cluster basis
\begin{eqnarray}
  \mid \Psi_{d'}^{J=0,T=0}> & = & {\cal A} \; {\Biggl \vert}
  \Biggl [ \Bigl [ \Phi_{T}^{S_T=1,T_T=0} 
  (\b{\rho}_{T},\b{\lambda}_{T},\b{\eta}_T) \times  [211]^{C_T=1}_{{\bf 3}} 
\nonumber \\
 &\otimes & \Phi_{D}^{S_D=0,T_D=0} (\b{\rho}_{D}) 
  \times [11]^{C_D={\bar 1}}_{{\bf {\bar 3}}} \Bigr ]^{S=1,T=0}
  \otimes\chi_{L=1}({\bf R}) \Biggr ]^{J=0,T=0} \ [222]^{C=0}_{{\bf 1}}  
\Bigr >, 
\label{rgmwf}
\end{eqnarray}
where $\Phi_T^{S_T=1,T_T=0}(\b{\rho}_{T},\b{\lambda}_{T},\b{\eta}_T)$
and $\Phi_D^{S_D=0,T_D=0}(\b{\rho}_{D})$ are the internal wave functions 
of the tetraquark (T) and diquark (D) clusters, respectively, and
$\chi_{L=1}({\bf R})$ is the yet unknown relative wave function of the
two colored clusters, projected onto good angular momentum L=1. 
The Jacobi coordinates introduced in 
Eq.\ (\ref{rgmwf}) are depicted in Fig.\ \ref{fig4}.
As usual in Resonating Group Method (RGM) \cite{wil77} calculations, 
ground state $s^2$  and $s^4$  harmonic oscillator wave functions 
are used for the diquark and tetraquark clusters, respectively.

Eq.\ (\ref{rgmwf}) shows that the color triplet $({\bf 3})$
tetraquark with mixed permutational symmetry [211], 
and the color antitriplet $(\bar {\bf 3})$ diquark with permutational
symmetry [11] is coupled to an overall color-singlet (${\bf 1}$) 
six-quark state with mixed permutational symmetry [222] in color space.
In contrast to mesons and baryons, the color state of a genuine $q^6$ system, 
is by itself not fully antisymmetric. There are both 
antisymmetric and symmetric quark pairs. Thus, in a compact 
$q^6$ system, there is no factorization of the color space 
and the flavor-spin-orbital space, i.e., the color dynamics and the 
orbital structure of the system are correlated in a more complex manner
\cite{Lip73}.  
This can be seen as an indication that the transition between $q^3$ and
compact $q^6$ systems is not as trivial as usually assumed. 

We recall that the stringlike bag model employs impenetrable diquark 
and tetraquark clusters, and the clusters cannot merge into a single bag.  
In contrast, an RGM calculation allows for a continuous transition from 
the compound $q^6$ state, where all quarks are in a single potential well, 
to the $q^2-q^4$ clusterized state, where one has two clearly separated 
bags. There is no artificial boundary 
between these extreme configurations, and they are both described by one 
and the same Resonating Group Method wave function.
Furthermore, the $q^3-q^3$ and $q^5-q^1$ partitioning into colored
clusters, as well as the $q^3-q^3$ split into color-singlet clusters 
is automatically included in the present theory.
These important properties are a consequence of the Pauli principle
on the quark level which is ensured by the antisymmetrizer ${\cal A}$
\begin{equation}
  {\cal A}= 1 - 8P_{46}^{OSTC} +6P_{35}^{OSTC} P_{46}^{OSTC}\; ,
\label{ant}
\end{equation}
where $P_{ij}^{OSTC}$ is the permutation operator of 
the i-th and j-th quark in orbital (O), spin-isospin (ST) and color space (C). 
The direct, one-quark- and two-quark-exchange contributions associated with 
the three terms in the antisymmetrizer of Eq.\ ({\ref{ant}) are shown
for the case of the one-gluon-exchange potential of Eq.\ (\ref{gluon}) 
in Fig.\ \ref{fig5}.

The solution for the unknown relative wave function $\chi_L({\bf R})$ 
and the unknown eigenenergy $E=M_{d'}$ 
is obtained from the Ritz variational principle
\begin{equation}
  \delta\left[{ \langle\Psi_{d'}\vert H-E\vert \Psi_{d'} \rangle
  \over\langle\Psi_{d'}\vert \Psi_{d'}\rangle  }\right ]=0,
\label{Ritz}
\end{equation}
where the variation is performed with respect to the relative wave function
$\chi_L({\bf R})$. 
After expanding the relative motion wave function $\chi_L({\bf R})$ in a 
finite sum of Gaussians centered around the generator coordinates ${\bf s}_i$
(Generator Coordinate Method (GCM)) 
\begin{equation}
  \chi_L ({\bf R}) =  \hat P_L \;  
  \sum_i C_i\; \bigg(\frac{4}{3\pi b_6^2}\bigg)^{3/4} \;
  {\exp}\left (-{2 \over 3}({\bf R} -{\bf s}_i)^2/{b_6^2} \right ) 
  \quad \Longrightarrow \quad 
  \vert \Psi_{d'}\rangle =: 
        {\cal{A}} \sum_i C_i\;\vert {\rm{GCM}}_i ({\bf s}_i) \rangle 
\label{relgauss} 
\end{equation}
and projecting 
onto good angular momentum $L$ via 
$\hat P_L \equiv 1/\sqrt{4\pi} \; \int d\hat s_i \cdot Y^L(\hat s_i)$ 
(cf.\ the appendix for more details),  
Eq.\ (\ref{Ritz}) transforms into a generalized algebraic eigenvalue problem 
\begin{equation}
  \sum_i {\cal{H}}_{ij} (s_i,s_j) \; C_i =E\;\sum_i {\cal{N}}_{ij} (s_i,s_j) 
  \; C_i \quad .
\label{rgmgcm}
\end{equation}
The analytic expressions for the Hamiltonian kernel 
${\cal{H}}_{ij} (s_i,s_j) = 
   \langle {\rm{GCM}}_i\vert {\cal{A}} H\vert {\rm{GCM}}_j\rangle$ 
and the norm kernel 
${\cal{N}}_{ij} (s_i,s_j) =  
   \langle {\rm{GCM}}_i\vert {\cal{A}} \vert {\rm{GCM}}_j\rangle$ 
of Eq.\ (\ref{rgmgcm}) can be found in the appendix. 
The norm matrix ${\cal{N}}_{ij} (s_i,s_j)$ on the r.h.s.\ of Eq.\ 
(\ref{rgmgcm}) reflects the non-orthogonality of the Gaussian basis functions
in Eq.\ (\ref{relgauss}).
 For simplicity, and to avoid a proliferation of parameters, the
same variational parameter $b_6$ for the internal and relative 
motion wave functions of the six-quark system is used.
This greatly facilitates the calculation
of the norm  and Hamiltonian integrals and restricts 
the variational space spanned by Eq.\ (\ref{relgauss}) only slightly.

Having solved the generalized eigenvalue problem, we determine
the mass of the $d'$ and the optimal harmonic oscillator 
parameter $b_6$, which plays the role of a nonlinear variational 
parameter, by minimization of the 
$d'$ mass:
\be
\label{mini}
\partial E_{\rm{d'}}(b_6) / \partial b_6 =0.
\ee
We emphasize that the $b_6$ values obtained in this way
are some 20-30$\%$ larger than the corresponding values $b_3$ which 
minimize the nucleon mass. This observation is the basis for
our conjecture that the two-body confinement strength in a
compound six-quark system such as the $d'$  is weaker than in 
a three-quark system (see Sect. \ref{subsec:weak}).

\section{Results and Discussion}   
\label{sec:res}

In this section we present the results of an RGM calculation for
the mass and size of the $d'$ dibaryon and compare it
with the corresponding shell model and stretched bag model results 
\cite{buc95}. Because  a corresponding 
RGM calculation for the deuteron,
based on the same Hamiltonian and colorless nucleon cluster wave functions
accurately reproduces the deuteron energy and wave function \cite{val95}, 
we are confident that the absolute mass scale of our predictions is correct, 
and the calculated $d'$ masses are reliable.

\subsection{Mass of the $d'$} 
\label{subsec:mass} 

The mass of the $d'$ is obtained as the lowest eigenvalue $E$ of the bound 
state GCM equation (\ref{rgmgcm}), and by subsequent optimization 
with respect to $b_6$ according to Eq.(\ref{mini}). It is interesting to 
isolate the effect of the Pauli principle on the mass of the $d'$ by 
performing  a calculation where the quark exchange kernels have been turned 
off, i.e. only the first term in the antisymmetrizer of Eq.(\ref{ant}) is 
taken into account, before solving Eq.(\ref{rgmgcm}). We use the same $b_6$ 
as in the full calculation including all exchange kernels.
Results for $M_{d'}$ and $b_6$ are shown in Table \ref{table:gcmmass}
for the parameter sets of Table \ref{table:para}. 
A comparison of the results for the $d'$ mass with 
and without quark exchange show that the quark exchange interactions 
contribute an                 
additional energy of 50--120 MeV depending on the model of
confinement (see sets I-VI in Table \ref{table:gcmmass}).

We note that the $d'$ mass calculated with model I  
(quadratic confinement and one-gluon exchange) 
is some 550 MeV above the experimental mass of the $d'$.
An RGM calculation without $\pi$- and $\sigma$-meson exchange 
forces in the Hamiltonian, but with a $q^3$-$q^3$ clusterization
of the quarks has already been performed many years ago \cite{cve80}.
The quark-quark interaction of Ref. \cite{cve80} is very similar
to model I, and the method
of fitting the quark model parameters to single-baryon properties 
is the same as in the present work.  
Although the authors of Ref. \cite{cve80} use two color-octet 
three-quark clusters, i.e., 
$({\bf 8} \otimes {\bf 8})^1$ 
whereas we use color-(anti)triplet diquark and tetraquark clusters, i.e., 
$({\bar {\bf 3}} \otimes {\bf 3})^1$,
it is nevertheless meaningful to compare their and our results. 

Due to the quark antisymmetrizer 
${\cal{A}}$ of Eq.\ (\ref{ant}) acting on all six quarks, the three 
different arrangements of the six-quarks into {\it colored} clusters, namely 
$q^3$-$q^3$, $q^2$-$q^4$, and $q^1$-$q^5$ are approximately equivalent for 
the relevant intercluster distances involved. In fact, in the region of 
complete cluster overlap, that is, in the limit for $s_i=s_j \to 0$ all 
three partitions have the same six-quark shell model limit, i.e. the unique
$s^5p^1$ TISM state. Thus, by virtue of the antisymmetrizer, the
RGM wave function not only contains the
 $({\bar {\bf 3}} \otimes {\bf 3})^1$ component, but also
the $({\bf 8} \otimes {\bf 8})^1$ component.
It is reassuring that the results of Ref. \cite{cve80}
$M_{d'}$=2540 MeV, and of the present work   
$M_{d'}$=2610 MeV (see set I in Table \ref{table:gcmmass})
are in very good agreement.
However, both are $\sim 500$ MeV above the experimental resonance position.
Before drawing any conclusions concerning the existence of 
the $d'$, we first study how these results depend on the choice
of the quark-quark interaction. 

In model II, we use the full Hamiltonian including one-gluon exchange, 
$\pi$- and $\sigma$-exchange \cite{com1}, 
and a quadratic confinement potential
between the constituent quarks.
The inclusion of the chiral $\pi$- and $\sigma$-exchange interactions reduces
the $d'$ mass by 160 MeV compared to model I. 
It is, however, with 2440 MeV still about 380 MeV higher than the 
experimental value.

Model III is very different from all other confinement models.
It uses a different confinement strength for the three-quark 
(nucleon) and the six-quark system ($d'$) and 
leads to a $d'$ mass close to the experimental result.
Before we try to justify this apparently `adhoc' step,
we study the dependence of the $d'$ mass on the radial
form of the confinement potentials given by 
Eqs.\ (\ref{linconf} -- \ref{erfconf}) and shown in Fig.3. 

The slope of models (IV-V) is for $r \ge 1$ fm 
considerably flatter than that of the quadratic confinement of 
Eq.\ (\ref{quaconf}). For these potentials, the stability 
condition Eq.(\ref{mini}) leads to larger oscillator parameters $b_6$
(larger dibaryons), which in turn results in a reduction of the kinetic 
energy and the $d'$ mass.   
For example, the linear confinement potential (model IV) 
reduces the $d'$ mass by about 90 MeV in comparison to the quadratic 
confinement (model II). For the $r^{2/3}$ confinement (model V), 
the corresponding reduction is 130 MeV.

Model VI, the color-screened error-function confinement, displays a 
qualitatively different 
behavior at small and large interquark distances. At short distances,
it is approximately linear and at larger distances it goes to a constant. 
While baryons feel mainly the linear rising part, 
dibaryons are due to their larger size 
also very sensitive to the long-range tail of the confinement potential.
The screening of color charges at larger interquark distances 
leads to a reduction of  
the $d'$ mass by 150 MeV compared to the quadratic confinement (model II) 
and also to a larger value for the oscillator parameter 
$b_6$.\footnote{In the suggested range for the color-screening length  
$0.8 {\rm fm} < 1/\mu < 1.2 {\rm fm}$ \cite{zha93}, our results are not very 
sensitive to the actual value of $1/\mu$.} However, the $d'$ mass is with 
2288 MeV still some 220 MeV above the experimental value of 
$M_{d'}^{exp}=2065$ MeV. If we use the $1-\exp(-\mu r^2)$ 
color-screening model of Ref. \cite{Wan92} with $\mu= 1$ fm$^{-2}$
for all quark pairs in the system, we obtain $ M_{d'}=2468$ MeV at 
$b_6=0.82$ fm.
In this context, we mention that at interquark distances  
corresponding to 3-4 times the screening length 1/$\mu$= 0.8 fm,
the effective confinement force 
$- \partial V^{\rm{conf}}_{\rm{erf}} (r) / \partial r$ 
between the colored clusters tends quickly to zero.
The stability of the system is nevertheless preserved 
due to Eq.(\ref{mini}) and our use of Gaussian trial 
wave functions. Model VI represents a 
limiting confinement model especially for dibaryons, and we expect 
that the 
$d'$ mass cannot be further reduced by an significant amount by considering
another radial form of the confinement potential. 

In Table \ref{table:contributions} we list 
the individual contributions of the kinetic, confinement, 
one-gluon-, one-pion-, and one-sigma-exchange potentials to the $d'$ mass 
for the different confinement models.
The results in parentheses give the corresponding contributions 
without the quark exchange diagrams. One observes a drastic reduction of the 
kinetic energy for the larger dibaryon systems. We also see that 
the dominant attraction is due to the color-Coulomb part of the interaction.

Thus, we observe a clear trend when going from a quadratic (model II) to a 
color-screened confinement model (VI): 
The slower the increase of the confinement 
potential at larger $r$, the larger the size $b_6$, and the 
smaller the mass of the $d'$. However, even for the 
limiting case of the color-screened confinement potential, the results for 
the $d'$ mass are 220 MeV above the empirical $d'$ mass.
We conclude that in the present constituent quark model framework, 
a change in the radial form of the confinement potential is insufficent 
to generate a $d'$-dibaryon with a mass of 2065 MeV.

\subsection{Wave function and size of the $d'$} 

\label{subsec:size} 

We turn now to the discussion of the relative wave function 
as obtained from eigenvalue equation (\ref{rgmgcm})
and the radius of the $d'$. 
Fig.\ \ref{fig6} shows the wave function $\chi_{L=1}(R)$ for the two extreme 
confinement models II and VI, using the same Hamiltonian as in the 
three-quark sector. The RGM wave function is calculated both 
with (Pauli-on) and without (Pauli-off) the quark exchange diagrams. 
For the quadratic confinement model II, one observes that
the relative wave function has already died out at intercluster distances 
of about  2.5 fm, while for confinement model VI
it still has an appreciable amplitude even at 4 fm. 
The quark exchange diagrams 
lead for model II to an additional attraction between the 
two colored clusters, as reflected
by the smaller radial extension of the relative wave function,
while they are rather unimportant for model VI.

A quantitative measure of the extension of the relative cluster 
wave function is the root mean square (rms) distance, $R_{d'}^{RGM}$,
which measures the mean distance between the $q^2$ and $q^4$ clusters
\be
\label{rgmrad}
(R_{d'}^{RGM})^2 =
\frac{\langle \chi_{L=1}(R)\vert R^2\vert \chi_{L=1}(R)\rangle }  
             {\langle \chi_{L=1}(R)\vert \chi_{L=1}(R)\rangle }.
\ee
Here, $R$ is the relative coordinate connecting the centers of the two 
clusters. This distance corresponds to the length $l$ 
of the color flux tube in the stringlike bag  model. 

The rms distance of Eq.(\ref{rgmrad}) 
does not yet take into account the finite size of the clusters. 
In the present model these are given as
\be
\label{diquarkrad}
r_D^2=
\left \langle \Phi_D \left \vert {1 \over 2} \sum_{i=1}^2 
({\bf r}_i-{\bf R}_D)^2 
\right \vert \Phi_D \right \rangle =  {3 \over 4} \, b^2; \qquad 
r_T^2=\left \langle \Phi_T \left \vert {1\over 4} 
\sum_{i=1}^4 ({\bf r}_i-{\bf R}_T)^2  \right \vert \Phi_T \right \rangle  =    
{9 \over 8} \, b^2.
\ee
Here, ${\bf R}_D$ and ${\bf R}_T$ are the center of mass coordinates of 
the diquark and tetraquark, respectively. 
One then finds for the total $d'$ radius in the RGM formalism 

\be
\label{totrad}
r_{d'}^2= r_D^2 + r_T^2 + ({1 \over 2} R_{d'}^{RGM})^2 + r_q^2,
\ee
where the last term is due to the finite size of the constituent quark.
In Table \ref{table:gcmrad}, the radii of the diquark, tetraquark, and the 
relative cluster wave function, $R_{d'}^{RGM}$ (with
are shown for different confinement potentials. 
Neglecting quark exchange between clusters 
increases $R_{d'}^{RGM}$ typically by about $10\%$ (numbers in parantheses).
This shows that the quark exchange matrix elements  
provide additional color attraction. 
Depending on the confinement model, 
the rms distances $R_{d'}^{RGM}$ lie between 1.0 and 2.0 fm,
while the corresponding tetraquark radii, $r_T$, vary between 
0.74 fm and 1.15 fm.

In order to have
another measure of the size of the $d'$, we calculate the rms distance 
between an arbitrary pair of quarks in the 
dibaryon\footnote{Note that in the case of the nucleon the mean distance 
between any pair of quarks is $b_3 \cdot\sqrt{3}\simeq 1 {\rm{fm}}$.}

\be
\label{rmsdis}
(r_{qq}^{RGM})^2=
    \left \langle \Psi_{\rm{d'}} \left \vert 
         \frac{1}{N_{\rm{pairs}}} \sum_{i<j}^6
         ({\bf r}_i -{\bf r}_j)^2
         \right \vert \Psi_{\rm{d'}} \right \rangle 
\ee 
where $N_{pairs}=15$ is the number of quark pairs 
in a six-quark system. The spatial extension of the 
quark distribution in the $d'$ is close to the value given by 
Eq.(\ref{rmsdis}). The radius defined in 
Eq.(\ref{rmsdis}) facilitates the comparison with 
the shell model description  of the $d'$ 
\cite{glo94,wag95,buc95}.
The numerical results for $r_{qq}^{RGM}$ 
are given in the second to last column of Table \ref{table:gcmrad}.

\subsubsection{Comparison with the six-quark shell model} 

\label{subsubsec:compshell} 

In Fig. \ref{fig6} we also compare the $q^2-q^4$ relative RGM wave function 
with an unperturbed six-quark $1\hbar\omega$ ($s^5p^1$) harmonic 
oscillator wave function 
\begin{equation}
  \chi^{{s^5p^1}}_{\rm{L=1}}(R) = 
  \frac{16}{3\sqrt{3\sqrt{3\pi } b_6^5}}\;
   R \  {\exp} \left ( {-\frac{2}{3}\frac{R^2}{b_6^2}} \right )\; ,
\label{ho}
\end{equation}
with the same oscillator parameter $b_6$.
Our previous results 
\cite{wag95} using the translationally invariant shell model (TISM) 
have shown that admixtures of excited shell model states to the unique lowest 
lying $s^5p^1$, 1$\hbar\omega$ shell model state of Eq.(\ref{ho}) are small.

For the quadratic confinement model II, 
the $d'$ wave function does not display a 
pronounced clusterization but resembles rather closely the six-quark harmonic
oscillator wave function of Eq.(\ref{ho}).
The agreement between the pure harmonic oscillator
wave function and the RGM wave function is
more complete when the quark exchange diagrams are included.
On the other hand, the RGM wave function of model VI extends 
to larger intercluster distances
$R$ than the unperturbed shell model state of Eq.(\ref{ho}). This 
is a consequence of the color-screened confinement potential,
which provides only weak confinement forces at large interquark distances.  
In order to describe the long-range RGM wave function 
in the shell model, one would need substantial admixtures of excited 
harmonic oscillator states.
As one might expect, the effect of the quark exchange diagrams
on the wave function 
is here much smaller than for the quadratic confinement case, 
which can also be seen from Table \ref{table:gcmmass}. 

For the wave function of Eq.\ (\ref{ho}), the mean square (ms) 
distance between the clusters is given by 
\be
\label{horad}
(R_{d'}^{HO})^2=\left \langle  \chi^{{s^5p^1}}_{\rm{L=1}} \left  \vert
R^2 \right \vert  \chi^{{s^5p^1}}_{\rm{L=1}} \right \rangle = 
{15 \over 8} \, b_6^2.
\ee
For all but one model, the rms distance between the $q^2$ and $q^4$ 
clusters as calculated in RGM 
is slighly larger than the corresponding quantity in the harmonic oscillator
model. Only in the case of the color-screened confinement 
potential (model VI), 
we find a pronounced increase of $R_{d'}^{RGM}$ by $40\%$ 
compared to $R_{d'}^{HO}$, 
indicating that the system is partially clusterized. 

In the case of the TISM wave function including shell model 
configurations up to 
$3 \hbar \omega$ \cite{wag95}, the rms-distance between any two quarks in 
the $d'$ is given by
\be
(r_{qq}^{TISM})^2 = 
b_6^2 \, \left (\frac{17}{5}+(1-\alpha^2)\frac{4}{5} \right ),
\ee
where $\alpha^2$ denotes the probability of finding the (unique) 
energetically lowest lying N=1 shell model
configuration in the total TISM wave function of the $d'$. 
The values given in Table \ref{table:gcmrad} are obtained for a harmonic
oscillator parameter $b_6$, which minimizes the $d'$ mass 
(see Table \ref{table:gcmmass}).

For models I -- V, the rms distance between an arbitray pair of quarks 
$r_{qq}^{RGM}$ 
follows closely the corresponding TISM result 
$r_{qq}^{TISM}$. 
Only for the color-screened confinement (set VI), the average distance between
quarks in the cluster model $r_{qq}^{RGM}$ 
is increased by about 10$\%$ with respect to the corresponding 
shell model result $r_{qq}^{TISM}$.  This is mainly due to the 
{\it intercluster} quark pairs, 
which due to the screening are preferably at larger relative distances.

Thus, a comparison of the cluster model and shell model results
\cite{wag95} 
reveals an overall agreement for the orbital structure and size of the $d'$. 
In addition, the two calculations agree also 
for the individual kinetic and different potential energy contributions 
to the $d'$ mass (as given for the cluster model
 in Table \ref{table:contributions}), 
and thus also for the $d'$ mass itself. 
The agreement is quantitative for models I-V.
However, the longer tail of the RGM wave function of model VI
shown in Fig.\ \ref{fig6} and the larger rms-radius listed in 
Table \ref{table:gcmrad} show that certain confinement models
lead to deviations from a pure $s^5p^1$ shell model structure 
and favor a clusterization of the quarks into colored quark clusters. 

Let us briefly discuss the reasons for the quantitative agreement 
between the two structurally quite different calculations. 
The outer product of the $[4]_O$ (tetraquark) and $[2]_O$ (diquark) orbital 
symmetries ($s^4$ and $s^2$) gives according to Littlewood's theorem 
the following  $S_6$ permutational symmetries in orbital space
\begin{equation}
\label{sym}
  [4]_O \otimes [2]_O = [42]_O \oplus [51]_O \oplus [6]_O.
\end{equation}
For $d'$ quantum numbers, the fully symmetric $[6]_O$ orbital symmetry, 
corresponds to a spurious center of mass  excitation of the six-quark state,
and is automatically excluded in both approaches. 
The orbital symmetries $[51]_O$ and $[42]_O$ in Eq.(ref{sym}) are
also included in the enlarged N=3$\hbar\omega$ shell model basis 
\cite{wag95}.  
Analogously, the outer product of the two clusters in spin-isospin space 
leads to
\begin{equation}
\label{STSYM}
  [31]_{ST} \otimes [2]_{ST} = [51]_{ST} \oplus [42]_{ST} 
            \oplus [33]_{ST} \oplus [411]_{ST} \oplus [321]_{ST} \; .
\end{equation}
Comparison with Eq.\ (10) in our previous TISM calculation \cite{glo94} 
shows that the 
$q^2-q^4$ cluster model wave function comprises the same $S_6$-symmetries 
in spin-isospin space (with the exception of the $[2211]_{ST}$ symmetry
which does not appear in the cluster model) as 
the enlarged shell model basis.  Thus the trial function space spanned by 
both sets of basis functions is nearly equivalent.

\subsubsection{Comparison with the stringlike bag model} 

\label{subsubsec:compbag} 

It is interesting to compare the description of the $d'$ in the cluster model
with the one in the stringlike bag model.  
With respect to the sizes, the cluster model values $R_{d'}^{RGM}$ and $r_T$
have to be compared to the length $l$ and radius $R_0$ of the color flux tube 
in the stringlike bag model. 
In the stretched  bag model \cite{mul80}, the radius $R_0$ and length $l$ of 
the flux tube are given as 
$$R_0=\sqrt{{\alpha \over 24 \pi B} }, 
\qquad  l={M_{d'} \over 4 \pi B \, R_0^2 } $$ 
respectively,
where $\alpha=1.1$ GeV$^2$ is the ``universal'' string tension, 
$B=59 $ MeV fm$^{-3}$ the bag constant, and $M_{d'}=2100$ MeV is 
the mass of the resonance. One then obtains for the ``universal'' 
radius of the color flux tube $R_0=1.1$ fm and
$l=2.3$ fm \cite{mul80}. 
 
According to Ref. \cite{mul80}, the stretched bag model is only valid if 
$l > 2 R_0$, and for a low-lying resonance of $M_{d'}=2100$ MeV this 
condition is barely satisfied. If we now identify the bag model quantities 
$l$ and $R_0$  with the cluster model quantities $R_{d'}^{RGM}$ and $r_T$ 
respectively, we see that this condition is generally not satisfied in 
the present calculation.
In fact, from table \ref{table:gcmrad} it is evident, that the sum
of the cluster radii exceeds the intercluster radius by about half the
diquark radius. This means that there is considerable overlap between
the clusters. Only for the color-screened confinement potential, 
we observe a modest clusterization that is, however, 
insufficient to justify a stringlike bag model treatment.

At this point, we comment on the universality of the string 
tension $\alpha$ in the stringlike bag model \cite{mul80}, which 
predicts a series of orbitally excited meson, baryon, and dibaryon states 
with orbital angular momentum $L$, and
with masses given by 
\be
\label{Regge}
M^2 = \alpha \,  L + M_0^2. 
\ee
The mass of the clusters at the ends of the string, $M_0$, 
is calculated in the 
{\it spherical} bag model, and $\alpha$ is the common string-tension 
for mesons ($q-{\bar q}$), baryons ($q^1-q^2$), and dibaryons
($q^4-q^2$) which is given by:
\be
\label{stringtension}
\alpha =\sqrt{8 \pi B \alpha_s f_C^2},  
\ee
where $B$ is the bag constant determining the constant 
color-electric field strength ${\bf E}$ in the string, 
$\alpha_s$ is the quark-gluon coupling,
and $f_C^2$ is the eigenvalue of the quadratic Casimir operator of 
$SU(3)_{color}$ in a given color representation.  
It is important to recall that the string tension $\alpha$ and the 
related transverse radius of 
the string $R_0$ are defined only for well separated clusters with 
sufficiently high relative angular momentum $L$ \cite{mul80}. 

A universal string tension $\alpha$ results because it is {\it assumed } 
that the individual clusters are so far apart that they are always 
in the color-triplet ${\bf 3}$ or color-antitriplet ${\bar {\bf 3}}$ 
representation, for which the Casimir operator has the common 
eigenvalue $f_C^2=4/3$. In this case, there is only one type of color string 
with a unique string tension given by Eq.(\ref{stringtension}). 
Obviously, this assumption is justified as long as the clusters do not 
overlap. However, we have seen that for 
most confinement models there is considerable overlap between 
the colored diquark and tetraquark clusters.
Consequently, the tunneling of quarks from one cluster to the other can 
change the color-representation 
(e.g. into ${\bf 8} \otimes \bar{ {\bf 8}} $, or 
${\bf 6} \otimes \bar{ {\bf 6}} $) of the clusters and therefore also the 
string tension. For the octet and sextet color representations, which are 
not present in mesons and baryons, one has $f_C^2=3$  and  $f_C^2=10/3$ 
respectively. Thus, one obtains a stronger string tension in both cases. The 
authors of Ref. \cite{mul80} only include the energetically most favorable 
case of a ${\bf 3} \otimes \bar{ {\bf 3}} $ string. 
If the Pauli principle is to be rigorously satisfied in Ref.\cite{mul80} also
the ${\bf 6} \otimes \bar{ {\bf 6}} $ and the ${\bf 8} \otimes {\bf 8} $ 
color configurations would have to be admixed and the corresponding 
$d'$ mass would come out higher. This qualitatively agrees with our
results for the $d'$ mass 
(see Tables \ref{table:gcmmass}-\ref{table:contributions} ) 
which are increased by the quark exchange interactions.

There is another difference between the present RGM calculation 
and the stringlike bag model. For the meson ($q-{\bar q}$), 
baryon ($q-q^2$), and dibaryon ($q^4-q^2$) resonances, the sizes of the 
color sources (clusters) and therefore the transverse radii of the flux tubes 
are in reality rather different. A tetraquark in a dibaryon has a larger 
radius than a single quark in a baryon. Although a tetraquark and a single 
quark carry the same color charge, the energy density of the color electric 
field, $E^2/2$, in the neighborhood of an extended tetraquark is smaller than 
in the neighborhood of a single quark. This difference is rather important 
for low angular momentum states. In the stretched bag model, 
the energy density of the string is given by the bag 
constant $B$. A decrease of the energy density of the string 
for $q^4-q^2$ dibaryons compared to $q-q^2$ baryons is related to a decrease 
of the effective bag constant $B$ 
\footnote{The universality of the bag constant 
for baryons and dibaryons has been questioned before $B$ \cite{kon86}.}.
According to Eq.(\ref{stringtension}) 
this leads to a reduced effective string tension for dibaryons. 
This reduction of the string tension clearly outweighs its increase due to 
quark exchange.

\par 
In conclusion, the assumption of a universal string tension is  
{\it not} justified for {\it ground states}, such as the pion, nucleon, 
or the $d'$ dibaryon. We emphasize once again that the stringlike bag model 
is valid only for {\it higher} angular momentum states, where the clusters 
are well separated due to the centrifugal barrier, and where the tunneling 
of quarks, as well as the size difference of the color sources can be 
neglected.

\subsection{The $d'$ as an indication for 
a weaker confinement strength in a compound six-quark system?} 
\label{subsec:weak} 

So far we have studied how different confinement models affect the mass 
and characteristic size of the dibaryon. In all cases, we obtained dibaryon 
masses some 200-400 MeV above the experimental value. 
We point out that up to this point, we have used the same
parameters as previously determined from the properties of color-singlet 
baryons (see Eq.(\ref{constraints})). 
Although it is reasonable to expect such a universality of parameters 
for the microscopically better founded
gluon-, pion- and sigma-exchange interactions, the 
assumption that the confinement strength remains unmodified when 
going from a three-quark to a six-quark system is most likely too restrictive. 

At present, there is no commonly accepted theory of color confinement. 
Different models of 
confinement have a limited range of validity. From the 
experimental 
information on the excited nucleon spectrum, combined with certain 
assumptions concerning the Lorentz structure, radial form, and 
color-dependence, phenomenologically successful confinement models for 
baryons have been constructed. However, it is not at all obvious that one can 
extrapolate the experience gathered in the color-singlet three-quark sector to 
compound six-quark systems. 
We recall that the confinement strength 
$a_c$ is usually determined from the experimental level-spacing $\omega$ 
between excited single-baryon states.
Unfortunately, there is hardly any three-star data 
on excited six-quark resonances that would shed 
light on the confinement dynamics in a six-quark system.  

We have seen that in the 
$d'$, the average distance between any two quarks is somewhat larger
than in the nucleon. Consequently, the $d'$ tests the confinement 
interaction at larger quark-quark distances where new physical phenomena 
come into play.
For example, at large interquark distances, quark-antiquark pair 
creation \cite{zha93,Wan92} leads to a 
screening of the original color charges. This color-screening 
effect can be simulated by a reduced effective confinement strength in 
the six-quark system as compared to the three-quark system.
 
Although we do not know how to calculate the effective 
confinement strength for three- and compound six-quarks systems 
from first principles, we can gain some qualitative understanding 
of the changing confinement dynamics using the 
harmonic oscillator model. 
If the effective quark-quark interaction were a pure harmonic oscillator
confinement force, there would be an inverse proportionality between the 
confinement strength $a_{c}^{(N)}$
\begin{equation}
        {a}_c^{(N)} = {1\over 2 m_q b_{(N)}^4 \, N } 
\label{confstr}
\end{equation}
and the fourth power of the harmonic oscillator parameter $b_{(N)}$.

In model II, we have used one and the same 
two-body confinement strength for three-quark and six-quark systems 
in the numerical calculation, {\it even though the variational principle
of Eq.(\ref{constraints}) and Eq.(\ref{mini})   
tells us that the characteristic sizes of the $d'$ $(b_6)$  
are about $30\%$ larger than those of the nucleon $(b_3)$ }. 
Thus, in a model where the two-body harmonic oscillator 
confinement is the only quark-quark interaction, there is 
a unique relation between the two-body harmonic confinement strength 
and the size of the system. We take this observation as the basis for our
conjecture that the {\it effective} two-body confinement strength 
could be weaker than in a three-quark system \cite{buc95}.
Although the actual dependence of the effective 
confinement strength on the size of the system 
may be somewhat different for more realistic confinement models, we expect 
that an inverse proportionality between the confinement strength 
and the size of the system remains.

Model III differs from model II only in the strength of the 
parameter $a^{(6)}_c$, for which we have taken the value 
$a^{(6)}_c=5.0$ MeV/fm$^2\approx a^{(3)}_c/3$ \cite{com2}. 
All other parameters in the Hamiltonian are identical to the ones of 
model II. The proposed weaker two-body confinement in a six-quark system 
leads to a larger $d'$ and  a mass that is close to the experimental $d'$ 
mass.  Conversely, one could take the empirical $d'$ mass 
as evidence for a weaker {\it effective} two-body 
confinement strength in a compound six-quark system because 
in the present model this is the only way to obtain the 
$d'$ resonance mass, which is needed to fit the DCX data.   

\subsection{Two-baryon vs. six-quark description of dibaryons}
\label{sec:bound}

The use of a reduced confinement strength in a compound six-quark
system, where all quarks are basically in a single bag, is presumably
not in conflict with our previous results for the deuteron. 
The deuteron is well decribed by an RGM wave function that is built
from two colorless three-quark clusters. The contribution of compound
six-quark states to the deuteron wave function is rather small.

Here, we would like to contrast the description 
of the $d'$ dibaryon as a compound six-quark state with the
successful picture (in terms of quarks)
of the deuteron as bound system of two color-singlet objects.
This comparison clearly shows that the orbital structure of these systems 
is determined by an interplay of the Pauli principle and  
different parts of the effective quark-quark interaction.

It has been known for more than a decade that quark exchange between 
nucleons (Pauli principle) together with the spin-dependent quark-quark 
interactions provide an effective short-range repulsion in the $NN$ system 
\cite{fae85}. The latter prevents appreciable cluster overlap and is therefore 
mainly responsible for the successful description of the deuteron 
as a system of two colorless nucleons with an average distance of
about $4$ fm between the nucleons.
 To explain this in terms of 
a six-quark shell model we form the outer product of two  $s^3$ nucleons.
One then obtains the following spatial permutational symmetries of a six-quark
system 
$$
  [3]_O \otimes [3]_O = [6]_O \oplus  [42]_O \oplus  [51]_O \oplus  [33]_O.  
$$
In a six-quark system with even orbital angular momentum, only the 
$[6]_O$ and $[42]_O$ symmetries occur. The important role of the orbital 
$[42]_O$ symmetry for the short-range repulsion in the $NN$ system has been 
elucidated in Ref. \cite{fae85}. In the shell model, which is appropriate 
for short $NN$ distances, the short-range repulsion is achieved through a 
destructive interference of the excited $s^4p^2$ six-quark state and the 
$s^6$ six-quark state. The spin-dependent quark-quark interactions lower the 
energy of the excited $s^4p^2$ state and raise the energy of the $s^6$ state. 
As a consequence, the excited $s^4p^2$ state is admixed with equal weight but 
opposite sign to the $s^6$ ground state so that it interferes destructively. 
This in turn leads to an almost complete cancellation of the wave function in 
the region of cluster overlap and consequently to the suppression  of the 
deuteron wave function at short distances \cite{obu87}.  Furthermore, 
explicit calculation shows that the deuteron wave function is insensitive 
to the details of the confinement mechanism.

On the other hand, in the case of the $NN$-decoupled $d'$-dibaryon, 
the radial form of the confinement potential crucially affects the mass and 
wave function of the system as we have demonstrated. The spin-dependent 
interactions are of minor importance. The latter point is reflected 
by the small admixtures of excited states to the lowest lying $s^5p^1$ shell 
model state. Because the Pauli principle prevents a clusterization of the 
six-quarks into two colorless nucleons, the energetically lowest 
configuration is a compound six-quark state. Thus, for both, 
the deuteron and the $d'$, the Pauli principle determines whether the 
residual spin-dependent or the confining interactions prevail, and hence 
determines the orbital structure of the system. 

\section{Summary}
\label{sec:sum}

\par
In the present work, we have calculated the mass and wave function 
of a J$^P$=0$^-$, T=0 six-quark system, called $d'$, 
in a colored diquark-tetraquark cluster model using the Resonating 
Group Method (RGM). This method determines the orbital configuration of the
six-quark system dynamically, i.e. according to the given Hamiltonian. 
Thus we may test the validity of the assumption that the $d'$ is a 
stretched diquark-tetraquark
system. In contrast to the stringlike bag model, which employs a single
non-antisymmetrized $q^2-q^4$ dumbbell-like configuration, the present 
RGM calculation also includes other clusterizations, such as the
$q^1-q^5$, $q^3-q^3$ and the single 
$q^6$ state. This remarkable property is a consequence of the proper 
antisymmetrization of the total six-quark RGM wave function. 

\par
A major purpose of this work has been to study the  effect of quark exchange 
interactions between the colored clusters on the mass and wave function of 
the $d'$. The quark exchange interactions between the colored diquark and 
tetraquark  clusters {\it increase} the $d'$ mass by some 100 MeV and 
slightly decrease the size of the $d'$ compared to a calculation without 
quark exchange.  Our results for the $d'$ mass and wave function are rather 
similar to a previous six-quark shell model 
calculation \cite{glo94,wag95,buc95}.
This means that for confinement models I-V, the $d'$ is in reality 
not a clusterized $q^4-q^2$ state, but better described as a single 
six-quark $s^5p^1$ shell model state, where all quarks move in a common 
potential well.
For this conclusion, it is crucial that the quark 
exchange diagrams are included in the cluster model.

\par	
In order to investigate how our results depend on the model of 
confinement, we have studied various, commonly used 
confinement potentials.
All models yield $d'$ masses substantially larger than the empirical value.
Depending on the confinement model 
we obtain $d'$ masses that lie roughly  200-400 MeV above the experimentally 
required resonance energy of $M_{\rm{d'}}$=2065 MeV.
Thus, we conclude that irrespective of the radial form of the 
confinement potential, we cannot describe 
a $J^P=0^-$ $T=0$ dibaryon with a mass of $2065$ MeV if we use 
the same confinement strength as in a three-quark system. Can one 
conclude that the $d'$ does not exist? 

\par
In a six-quark system, such as the $d'$, the confinement interaction 
is tested in a heretofore unexplored regime 
where new physical phenomena, such as color-screening due to quark pair 
creation and many-body confinement forces  
are expected to play an important role. These long-range 
effects are difficult to accomodate within the standard {\it two-body} 
confinement force model of Lipkin-type with a universal 
confinement strength for baryons and compact six-quark states. 
In order to {\it model} the complex color dynamics in a compound
six-quark system, we have proposed that the effective two-body 
confining strength in the $d'$ is weaker than in a single nucleon (model III).
We then obtain a $d'$ mass, $M_{d'}=2092$ MeV, which is compatible with the 
experimentally suggested resonance mass. In the absence of a solvable theory 
of confinement, our harmonic oscillator model III should be viewed 
{\it as an attempt to model the more complicated color dynamics in a genuine
six-quark system}. Our conjecture of a weaker confinement strength is 
based on the observation that the average distance between any pair of quarks
in the $d'$ is significantly larger than in the nucleon.
In this long-range regime the confinement interaction is poorly understood.
A harmonic oscillator confinement model clearly shows the inverse relation 
between the size of the system and the confinement strength.
If the $d'$ is experimentally confirmed, it would be an indication 
that the {\it effective two-body} confinement strength is weaker in a genuine 
six-quark system with the size of the $d'$.

It remains to be seen whether our hypothesis of a weaker confinement strength 
in a compound six-quark system affects other results 
in the $B=2$ sector such as $NN$ scattering phase shifts or the deuteron 
electromagnetic form factors \cite{shi89}. 
According to our experience, these observables are rather 
insensitive to the details of confinement, and we do not expect
any qualitative changes of previous results. Nevertheless, new 
calculations, e.g. of deuteron observables should be carried out
to explicitly check the consequences of our conjecture.
 
An important test of the $d'$ hypothesis is the simultaneous description
of the mass and the free pionic decay width 
$\Gamma_{\rm{d'}} \simeq 0.5 {\rm{MeV}}$ 
of the $d'$ with the same set of parameters. Model III  
gives a {\it free} $d'$ decay width of $\Gamma_{d'} \approx 0.3 $ MeV, 
which is in agreement with experimental result \cite{ito96}.
In the future, a detailed analysis of the $Nd'$ and $d'd'$ 
interactions similar to the calculation of the $NH$ \cite{sak92} 
and $HH$ \cite{sak97} interactions, as well as a calculation of 
the cross section for $d'N \to NNN$ should be done. 
This will provide a stringent test of whether the theoretical $d'$ radius 
$r_{d'}=1.53$ fm is consistent with the empirical `in medium' $d'$ decay width
$\Gamma_{medium} \approx 5 $ MeV.

On the experimental side, the search for a clear $d'$ signature in elementary 
reactions, such as  $pp \to pp \pi^+ \pi^-$ 
\cite{bro96} and $\gamma d \to pn \pi^0$  \cite{Fil92}, has begun. 
The experimental search for dibaryons provides a unique
chance for probing the largely unexplored and little understood 
phenomenon of color confinement at larger interquark distances.
In this regime our traditional confinement models are likely to break down. 
If the existence of narrow dibaryons 
is confirmed, it will also have important implications for deep-inelastic 
electron scattering off nuclei \cite{cio95}, for the equation of state of 
nuclear matter at higher densities, and the inner structure of neutron 
stars \cite{kri95}. 

\medskip
{\bf Acknowledgements:} 
The authors thank K.\ Tsushima for calculational help at 
the early stage of the project. Stimulating discussions on the 
confinement issue with H.\ Clement, M.\ Engelhardt, L.\ Ya.\ Glozman, 
M.\ I.\ Krivoruchenko, D.\ B.\ Lichtenberg, E.\ Lomon, 
I.\ T.\ Obukhovsky, and M.\ Rosina are acknowledged. 
This work is supported in parts by the DFG Graduiertenkolleg Mu705/3, 
BMBF 06 T\"{u} 746(2) and DFG postdoctoral fellowship Wa1147/1-1.

\newpage

\centerline{ {\bf Appendix} }

In this appendix, we present all norm  and Hamiltonian kernel matrix 
elements neccessary to diagonalize the algebraic generalized eigenvalue
problem of Eq.\ (\ref{rgmgcm}).
We calculate the Hamiltonian and norm kernels
\begin{equation}
  {\cal{H}}_{ij} (s_i,s_j) = 
   \langle {\rm{GCM}}_i\vert {\cal{A}} H\vert {\rm{GCM}}_j\rangle 
  \quad ; \quad 
  {\cal{N}}_{ij} (s_i,s_j) =  
   \langle {\rm{GCM}}_i\vert {\cal{A}} \vert {\rm{GCM}}_j\rangle , 
\label{hamnorm}
\end{equation} 
where the abbreviation 
$\vert {\rm{GCM}}_i\rangle$  introduced in Eq.\ (\ref{relgauss}) denotes
\begin{eqnarray}
  \vert {\rm{GCM}}_i\rangle &\equiv &
  \frac{1}{\sqrt{4\pi }} \big[ (1/\pi b_6^2)^{3/4} \big]^6 
  \int d\hat s_i  \cdot 
  \bigg[ Z^{S=1,T=0,C=0} \times Y^{L=1}(\hat s_i) \bigg]^{J=0,T=0,C=0}  
  \times
\nonumber \\
  & \times & 
  \bigg\{ \prod_{\alpha =1}^4 \exp \left ({-\frac{1}{2b_6^2}({\bf r}_\alpha -
          \frac{2}{3} \frac{{\bf s}_i}{2} )^2} \right )  \bigg\} 
  \bigg\{ \prod_{\beta =5}^6 \exp \left ( {-\frac{1}{2b_6^2}({\bf r}_\beta +
          \frac{4}{3} \frac{{\bf s}_i}{2} )^2} \right ) \bigg\} 
\nonumber \\
  Z^{S=1,T=0,C=0} &=& 
  \Bigl [ \Phi_{T}^{S_T=1,T_T=0} 
  \times  [211]^{C_T=1}_{{\bf 3}} \otimes
   \Phi_{D}^{S_D=0,T_D=0}  
  \times [11]^{C_D=1}_{{\bar {\bf 3}}} \Bigr ]^{S=1,T=0}\, 
[222]^{C=0}_{{\bf 1}}
\label{gcmdef}
\end{eqnarray} 
In Eq.\ (\ref{gcmdef}) $Z^{S=1,T=0,C=0}$ is the spin(S)-isospin(T)-color(C) 
wave function of the $d'$, where 
the diquark ($S_D=0$, $T_D=0$, $C_D={\bar 1} $) 
and the tetraquark ($S_T=1$, $T_T=0$ $C_T=1$) 
STC-wave functions are coupled to the total STC wave function of the $d'$,
with $S=1$, $T=0$, and total color $C=0$ of the color singlet state 
[222]$_{{\bf 1}}$.
The projection onto good angular momentum $L=1$ of the orbital part of 
the matrix elements of Eq.\ (\ref{hamnorm})
is done after the integration over all quark coordinates
shown in Fig.\ \ref{fig4}:  
$\prod_{i=1}^6 \int d{\bf r}_i$.
Introducing the notation $A=1/{(2b_6^2)}$, we obtain for the orbital 
integrals before projection, first for the norm matrix kernels 
${\cal{N}}_{ij}$:
\begin{eqnarray}
  {\cal{N}}^D_{ij} &=& 
     \exp \left ( -\frac{2A}{3}({\bf s}_i-{\bf s}_j)^2 \right )
\nonumber \\[0.15cm]
  {\cal{N}}^X_{ij} &=& 
     \exp \left ( 
{-\frac{2A}{3}({\bf s}_i^2-{\bf s}_j^2)} \right ) \,
          \exp \left ( {+\frac{A}{3}{\bf s}_i\cdot {\bf s}_j} \right )
\nonumber \\[0.15cm]
  {\cal{N}}^{XX}_{ij} &=& 
          \exp \left ( {-\frac{2A}{3}({\bf s}_i^2-{\bf s}_j^2)} \right ) \, 
          \exp \left ( {-\frac{2A}{3}{\bf s}_i\cdot {\bf s}_j} \right ) 
\label{eq:xnpnorm}
\end{eqnarray}
Here and in the following, D refers to the direct kernel, X represents the 
one-quark-exchange diagrams for quark 4 and 6, and XX refers to the 
two-quark-exchange diagrams of the pairs (34)$\leftrightarrow$(56) in orbital
space.
The subscripts in the kinetic energy and the
two-body interaction kernels denote the quark coordinate. 

The non-projected kernels of the kinetic energy operator
$T_i =\frac{p_i^2}{2m_q}=-\frac{1}{2m_q} {\b{\nabla}}^2_i$ 
read
\begin{eqnarray}
  T_1^D ({\bf s}_i,{\bf s}_j) &=& -\frac{1}{2m_q} 
    \big[ -3A +\frac{A^2}{9} ({\bf s}_i-{\bf s}_j)^2 \big] {\cal{N}}^D_{ij}
\nonumber \\
  T_6^D ({\bf s}_i,{\bf s}_j) &=& -\frac{1}{2m_q}
    \big[ -3A +\frac{4A^2}{9} ({\bf s}_i-{\bf s}_j)^2 \big] {\cal{N}}^D_{ij} 
\nonumber \\
  T_1^X ({\bf s}_i,{\bf s}_j) &=& -\frac{1}{2m_q} 
    \big[ -3A +\frac{A^2}{9} ({\bf s}_i-{\bf s}_j)^2 \big] {\cal{N}}^X_{ij}
\nonumber \\
  T_5^X ({\bf s}_i,{\bf s}_j) &=& -\frac{1}{2m_q}  
    \big[ -3A +\frac{4A^2}{9} ({\bf s}_i-{\bf s}_j)^2 \big] {\cal{N}}^X_{ij}
\nonumber \\
  T_4^X ({\bf s}_i,{\bf s}_j) &=& -\frac{1}{2m_q} 
    \big[ -3A +\frac{A^2}{9} ({\bf s}_i+2{\bf s}_j)^2 \big] {\cal{N}}^X_{ij}
\nonumber \\
  T_6^X ({\bf s}_i,{\bf s}_j) &=& -\frac{1}{2m_q} 
    \big[ -3A +\frac{A^2}{9} (2{\bf s}_i+{\bf s}_j)^2 \big] {\cal{N}}^X_{ij}
\nonumber \\
  T_1^{XX} ({\bf s}_i,{\bf s}_j) &=& -\frac{1}{2m_q} 
    \big[ -3A +\frac{A^2}{9} ({\bf s}_i-{\bf s}_j)^2 \big] {\cal{N}}^{XX}_{ij}
\nonumber \\
  T_4^{XX} ({\bf s}_i,{\bf s}_j) &=& -\frac{1}{2m_q} 
    \big[ -3A +\frac{A^2}{9} ({\bf s}_i+2{\bf s}_j)^2 \big] {\cal{N}}^{XX}_{ij}
\nonumber \\
  T_6^{XX} ({\bf s}_i,{\bf s}_j) &=& -\frac{1}{2m_q} 
    \big[ -3A +\frac{A^2}{9} (2{\bf s}_i+{\bf s}_j)^2 \big] {\cal{N}}^{XX}_{ij}
\label{eq:xnpkin}
\end{eqnarray}
where the norm kernels ${\cal{N}}_{ij}$ of Eq.\ (\ref{eq:xnpnorm})
have been factored out.
For the two-body potential kernels depicted in Fig. \ref{fig5} (for the case
of the one-gluon-exchange potential), we arrive after some straightforward
algebra at:
\begin{eqnarray}
  V_{12}^D ({\bf s}_i,{\bf s}_j)    &=&
  V_{56}^D ({\bf s}_i,{\bf s}_j)    =
  {\cal{N}}^D_{ij}  \bigg( \frac{A}{\pi} \bigg)^{\! 3/2} \; (4\pi ) \;
  \int_0^\infty dr\; r^2\, \exp({-Ar^2})\; V(r) 
\nonumber \\[0.1cm]
  V_{46}^D ({\bf s}_i,{\bf s}_j)    &=&
  {\cal{N}}^D_{ij}  \bigg( \frac{A}{\pi} \bigg)^{\! 3/2}\! (4\pi )^2 \,
   \exp \left ( {-\frac{A}{4} ({\bf s}_i+{\bf s}_j)^2} \right ) \times
\nonumber \\
  &\times &
  \sum_{l,m} {Y_m^l}^\ast (\hat {\bf s}_i) {Y_m^l} (\hat {\bf s}_j) 
 \int_0^\infty dr \, r^2\, \exp({-Ar^2}) \; V(r)\, i_l(Ars_i)\, i_l(Ars_j)
\nonumber \\
  V_{12}^X ({\bf s}_i,{\bf s}_j)    &=& 
  {\cal{N}}^X_{ij}  \bigg( \frac{A}{\pi} \bigg)^{\! 3/2} \; (4\pi ) \;
  \int_0^\infty dr\; r^2\, \exp( {-Ar^2} )\; V(r) 
\nonumber \\
  V_{34}^X ({\bf s}_i,{\bf s}_j)    &=& 
  V_{56}^X ({\bf s}_i,{\bf s}_j)    = 
  {\cal{N}}^X_{ij}  \bigg( \frac{A}{\pi} \bigg)^{\! 3/2} \; (4\pi ) \;
  \exp \left ( {-\frac{A}{4} {\bf s}_j^2}  \right )
  \int_0^\infty dr\; r^2\, \exp({-Ar^2})\; V(r) i_0(Ars_j)
\nonumber \\[0.1cm]
  V_{36}^X ({\bf s}_i,{\bf s}_j)    &=& 
  V_{45}^X ({\bf s}_i,{\bf s}_j)    = 
  {\cal{N}}^X_{ij}  \bigg( \frac{A}{\pi} \bigg)^{\! 3/2} \; (4\pi ) \;
  \exp \left ( {-\frac{A}{4} {\bf s}_i^2} \right )  
  \int_0^\infty dr\; r^2\, \exp(-Ar^2)\; V(r) i_0(Ars_i) \nonumber \\[0.10cm] 
& & \\[0.15cm]
  V_{46}^X ({\bf s}_i,{\bf s}_j)    &=& 
  {\cal{N}}^X_{ij}  \bigg( \frac{A}{\pi} \bigg)^{\! 3/2}\! (4\pi )^2 \,
  \exp \left ( {-\frac{A}{4} ({\bf s}_i-{\bf s}_j)^2} \right ) \times 
\nonumber \\
  &\times& \sum_{l,m} (-)^l\cdot 
  {Y_m^l}^\ast (\hat {\bf s}_i) {Y_m^l} (\hat {\bf s}_j) 
 \int_0^\infty dr \, r^2\, \exp({-Ar^2})\; V(r)\, i_l(Ars_i)\, i_l(Ars_j)
\nonumber \\
  V_{35}^X ({\bf s}_i,{\bf s}_j)    &=& 
  \big( {\cal{N}}^X_{ij}/{\cal{N}}^D_{ij}\big)  \cdot 
  V_{46}^D ({\bf s}_i,{\bf s}_j) 
\nonumber \\
  V_{12}^{XX} ({\bf s}_i,{\bf s}_j) &=& 
  V_{34}^{XX} ({\bf s}_i,{\bf s}_j) = V_{56}^{XX} ({\bf s}_i,{\bf s}_j) = 
  {\cal{N}}^{XX}_{ij}  \bigg( \frac{A}{\pi} \bigg)^{\! 3/2} \; (4\pi ) \;
  \int_0^\infty dr\; r^2\, \exp({-Ar^2})\; V(r) 
\nonumber \\[0.15cm]
  V_{23}^{XX} ({\bf s}_i,{\bf s}_j) &=& 
  {\cal{N}}^{XX}_{ij}  \bigg( \frac{A}{\pi} \bigg)^{\! 3/2} \; (4\pi ) \;
  \exp\left ({-\frac{A}{4} {\bf s}_j^2} \right )  
  \int_0^\infty dr\; r^2\, \exp({-Ar^2})\; V(r) i_0(Ars_j)
\nonumber \\
  V_{25}^{XX} ({\bf s}_i,{\bf s}_j) &=& 
  {\cal{N}}^{XX}_{ij}  \bigg( \frac{A}{\pi} \bigg)^{\! 3/2} \; (4\pi ) \;
  \exp\left ( {-\frac{A}{4} {\bf s}_i^2} \right )  
  \int_0^\infty dr\; r^2\, \exp({-Ar^2})\; V(r) i_0(Ars_i)
\nonumber \\[0.1cm]
  V_{46}^{XX} ({\bf s}_i,{\bf s}_j) &=& V_{45}^{XX} ({\bf s}_i,{\bf s}_j) =
  \big( {\cal{N}}^{XX}_{ij}/{\cal{N}}^{X}_{ij}\big) \cdot 
    V_{46}^X ({\bf s}_i,{\bf s}_j) \nonumber \\
\label{eq:potiallg} 
\end{eqnarray}

In order to project all kernels onto good angular momentum $L=1$ 
in the relative coordinate, we first expand 
the scalar products of  
the generator coordinates (${\bf s}_i\cdot {\bf s}_j$) appearing in the
exponential functions in terms of spherical harmonics 
\begin{equation}
  \exp \left ( {-\alpha A{\bf s}_i\cdot {\bf s}_j} \right ) = 
  4\pi\; \sum_l (-)^l \;\hat l \; i_l(\alpha As_is_j) Y^l(\hat {\bf s}_i) 
      \cdot Y^l(\hat {\bf s}_j) \qquad .
\label{eq:besselx}
\end{equation}
The projection may then be performed by application of
the integral operator
$\hat P_{\rm{GCM}}^{L=1} = \int d\hat {\bf s}_i d\hat {\bf s}_j 
Y^{L=1}({\bf s}_i) \, Y^{L=1}({\bf s}_j) $. 

In the present case ($L=1$, $S=1$, $T=0$) we obtain the following 
expressions for the projected norm kernels 
\begin{eqnarray}
  {\cal{N}}^{D,proj}_{ij} &=& 
     \frac{1}{\sqrt{3}}\;
\exp\left ( 
{-\frac{2A}{3}({\bf s}_i^2+{\bf s}_j^2)} \right )
     \,\cdot\, i_1\biggl ( \frac{4A}{3}s_is_j\biggr ) \cdot \langle STC \rangle
\nonumber \\[0.25cm]
  {\cal{N}}^{X,proj}_{ij} &=& 
     \frac{1}{\sqrt{3}}\;\exp\left ( {-\frac{2A}{3}({\bf s}_i^2+{\bf s}_j^2)}
\right ) \,\cdot\, i_1\biggl ( \frac{A}{3}s_is_j\biggr ) 
\cdot \langle STC\rangle
\nonumber \\[0.25cm]
  {\cal{N}}^{XX,proj}_{ij} &=& 
     -\frac{1}{\sqrt{3}}\;
\exp\left ( {-\frac{2A}{3}({\bf s}_i^2+{\bf s}_j^2)} \right ) 
     \,\cdot\, i_1\biggl ( \frac{2A}{3}s_is_j \biggr ) \cdot \langle STC\rangle
\label{eq:xnpnormproj}
\end{eqnarray}
where the short-hand notation $\langle STC \rangle$ has been introduced to
denote the reduced norm operator matrix elements 
$\langle\vert\vert        1          \vert\vert\rangle , 
 \langle\vert\vert \hat P_{46}^{STC} \vert\vert\rangle , {\rm{and}}
 \langle\vert\vert \hat P_{35}^{STC}\cdot \hat P_{46}^{STC} \vert\vert\rangle$ 
in spin-isospin-color space. Explicit expressions are give below. 

Some of the projected kinetic energy kernels
are simply obtained by the substitution 
${\cal{N}}_{ij} \rightarrow {\cal{N}}^{proj}_{ij}$.  
The components proportional to 
$\sim ({\bf s}_i\cdot {\bf s}_j)\cdot {\cal{N}}_{ij}$ are obtained 
after application
of the trick:
\begin{equation}   
  ({\bf s}_i\cdot {\bf s}_j)\cdot 
  \exp\left ({-A{\bf s}_i\cdot {\bf s}_j}\right ) =
  -\frac{1}{A} \bigg\vert\; \frac{\partial}{\partial x}
  \exp \left 
({-Ax{\bf s}_i\cdot {\bf s}_j} \right )\bigg\vert_{x=1} \quad ,\quad 
  \bigg\vert\; \frac{\partial}{\partial x} i_1\big( xy\big) \bigg\vert_{x=1}
  = yi_2\big( y\big) + i_1\big( y\big) \; . 
\label{eq:besseltrick}
\end{equation}        

The replacement 
${\cal{N}}_{ij} \rightarrow {\cal{N}}^{proj}_{ij}$ applies also for most 
of the projected two-body potential matrix elements, except for
\begin{eqnarray}
  V_{46}^{D,proj} &(&s_i, s_j) =   
  \frac{(4\pi )}{\sqrt{3}} \,\bigg( \frac{A}{\pi} \bigg)^{\! 3/2}
  \exp \left ( {-\frac{11A}{12} ({\bf s}_i^2+{\bf s}_j^2)} \right ) 
  \cdot\langle STC\rangle\times 
\nonumber \\
  &\times &  \sum_{l} \bigg( \frac{\partial}{\partial x} i_l\big( x\big) \bigg)
  \cdot\hat l^2
 \int_0^\infty dr \, r^2\, \exp({-Ar^2})\; V(r)\, i_l(Ars_i)\, i_l(Ars_j)
  \; ;\; x=\frac{5}{6}As_is_j
\nonumber \\
  V_{46}^{X,proj} &(&s_i, s_j) =
  \frac{(4\pi )}{\sqrt{3}} \,\bigg( \frac{A}{\pi} \bigg)^{\! 3/2}
  \exp \left ( {-\frac{11A}{12} ({\bf s}_i^2+{\bf s}_j^2)}  \right )
  \cdot\langle STC \rangle\times 
\nonumber \\
  &\times &  \sum_{l} 
  \biggl( \frac{\partial}{\partial x} i_l\big( x\big) \bigg)
  (-)^l \cdot\hat l^2
 \int_0^\infty dr \, r^2\, \exp({-Ar^2})\; V(r)\, i_l(Ars_i)\, i_l(Ars_j)
  \; ;\; x=\frac{5}{6}As_is_j
\nonumber \\
  V_{35}^{X,proj} &(& s_i, s_j) =
  -\frac{(4\pi )}{\sqrt{3}} \,\bigg( \frac{A}{\pi} \bigg)^{\! 3/2}
    \exp \left ( {-\frac{11A}{12} ({\bf s}_i^2+{\bf s}_j^2)} \right )
  \cdot\langle STC\rangle\times 
\nonumber \\
  &\times & \sum_{l} \bigg( \frac{\partial}{\partial x} i_l\big( x\big) \bigg)
  (-)^l \cdot\hat l^2
 \int_0^\infty dr \, r^2\, \exp({-Ar^2})\; V(r)\, i_l(Ars_i)\, i_l(Ars_j)
  \; ;\; x=\frac{1}{6}As_is_j
\nonumber \\[0.2cm]
  V_{46}^{XX,proj} &(&s_i, s_j) = V_{35}^{X,proj} (s_i, s_j) 
\label{eq:wwproject}
\end{eqnarray}

Finally, we explicitly give the spin-isospin-color $\langle STC \rangle$ 
factors for the different
matrix elements, first for the norm kernel
\begin{eqnarray}
  \langle {\rm{(S=1, L=1)J=0, T=0}}  &\vert &\!\vert   1\;
  \vert\vert {\rm{ (S=1, L=1)J=0, T=0}} \rangle 
  =  \sqrt{3}
\label{eq:normstcme} \\
  \langle {\rm{ (S=1, L=1)J=0, T=0}}  &\vert &\!\vert 
\hat P^{STC}_{46}\;
  \vert\vert {\rm{(S=1, L=1)J=0, T=0}} \rangle 
  =  0
\nonumber \\
  \langle {\rm{ (S=1, L=1)J=0, T=0}}  &\vert &\!\vert 
  \; \hat P^{STC}_{35}\hat P^{STC}_{46}\;
  \vert\vert {\rm{ (S=1, L=1)J=0, T=0}} \rangle 
  =  +\frac{\sqrt{3}}{4}.
\nonumber
\end{eqnarray}
The same $\langle STC \rangle$ matrix elements apply to the kinetic energy 
and for the one-sigma-exchange potential, because as the norm operator, 
these operators do not depend on isospin, spin, or color.
Table \ref{table:stcelemente}
collects all neccessary $\langle STC \rangle$ matrix 
elements for the remaining interactions. 
In the calculation of these matrix elements, we have made use of 
the well-known identities 
\begin{equation}
  \lilj = -\frac{2}{3} +2\hat P_{ij}^C
\quad ,\quad 
  \sisj = -1 +2\hat P_{ij}^S
\quad ,\quad 
  \titj = -1 +2\hat P_{ij}^T
\label{eq:opper}
\end{equation} 
and also of the fractional parentage decomposition of the tetraquark and
diquark wave functions to evaluate the quark exchange operator acting in the
different spaces.

\newpage
\begin{table}[htb]
\begin{center} 
\begin{tabular}{ l  r | l  l  l  l  } 
\rule[-0.5mm]{0mm}{6.5mm}Conf.\   & Set & 
  $b_3$ [fm] & $\alpha_s$ & $a_c^{(3)}$ & 
  $a_c^{(6)}$ \\[0.15cm] \hline
\rule[-0.5mm]{0mm}{6.5mm}Q    
  & I   & 0.603 & 1.540 & 24.94 MeV/fm$^2$ & $= a_c^{(3)}$       \\[0.05cm] 
Q & II  & 0.595 & 0.958 & 13.66 MeV/fm$^2$ & $= a_c^{(3)}$       \\[0.05cm]  
Q & III & 0.595 & 0.958 & 13.66 MeV/fm$^2$       & 5.00  MeV/fm$^2$ \\[0.05cm]
L & IV  & 0.609 & 1.060 & 22.03 MeV/fm           & $= a_c^{(3)}$  \\[0.05cm]  
R & V   & 0.617 & 1.122 & 26.12 MeV/fm$^{2/3}$   & $= a_c^{(3)}$  \\[0.05cm]  
E & VI  & 0.648 & 1.374 & 46.70 MeV     & $= a_c^{(3)}$      
\end{tabular}
  \caption[Quark model parameters]{Quark model parameters.
     Sets I -- III are obtained for the quadratic confinement potential (Q)
     of Eq.\ (7). 
     Set I: without $\pi$- and $\sigma$-exchange potentials. 
     Set II: with regularized $\pi$- and $\sigma$-meson exchange potentials.
     Set III: same
     as set II but with reduced confinement strength in the six-quark
     sector.
     Set IV: with linear confinement (L) of Eq.\ (8)
     instead of quadratic confinement.
     Set V: with $r^{2/3}$ confinement interaction (R) of Eq.\ (9).
     Set VI: with the color-screened error-function confinement (E) of Eq.\
     (10) using an inverse 
     screening length of 1/$\mu$= 0.8 fm. 
     In columns 5 and 6 we  list the confinement strength $a_c^{(3)}$ 
     as determined from the nucleon mass
     and the confinement strength
     $a_c^{(6)}$ employed for the calculation of the dibaryon mass 
     and wave function as discussed  in Section IV.} 
  \label{table:para}
\end{center} 
\end{table}

\begin{table}[htb]
\begin{center}
\begin{tabular}{ l  c | c  c  c  c  c }
\rule[-0.5mm]{0mm}{6.5mm}Conf.\  
  & Set & $M_{2q}$ & $M_{4q}$ & $M_{d'}$ & $b_6$ &
  $M_{\rm{no QEX}}$  \\[0.10cm]
  &  & [MeV] & [MeV] & [MeV] & [fm] & [MeV] 
\\[0.15cm] \hline
\rule[-0.5mm]{0mm}{6.5mm}Q 
  & I   & 636 & 1501 & 2610 & 0.70 & 2490  \\[0.05cm]
Q & II  & 643 & 1456 & 2440 & 0.75 & 2316  \\[0.05cm]    
Q & III & 621 & 1309 & 2092 & 0.95 & 2013  \\[0.05cm]
L & IV  & 650 & 1431 & 2354 & 0.86 & 2261  \\[0.05cm]
R & V   & 653 & 1419 & 2313 & 0.93 & 2234  \\[0.05cm]
E & VI  & 661 & 1422 & 2288 & 1.08 & 2235  \\[0.05cm]
\end{tabular}
\caption[masses]{The mass $M_{\rm{d'}}$ and characteristic 
          size $b_6$ of the $d'$ as obtained according to Eq.(18)  
          for the six parameter sets of Table I. 
          The diquark and tetraquark masses are also shown.
          The masses in the last column are obtained in a calculation
          without quark exchange between the diquark and the tetraquark 
          clusters. As in table I, the abbreviations  
          Q, L, R, and E identify the four different 
          confinement models used in this work. The experimental mass of the 
          $d'$ is $M_{d'}= 2065 \pm 5 $ MeV [3]. }
\label{table:gcmmass}
\end{center}
\end{table}

\begin{table}[htb]
\begin{center}
\begin{tabular}{ l  c | c  c  c  c  c  c }
\rule[-0.5mm]{0mm}{6.5mm}Conf.\  & Set & 
  $E_{\rm{kin}}$ & $V_{\rm{conf}}$ &
  $V_{\rm{gluon}}$ & $V_\pi$ & $V_\sigma$ & $M_{d'}$  
\\[0.10cm]
  &  & [MeV] & [MeV] & [MeV] & [MeV] & [MeV] & [MeV]         
\\[0.15cm] \hline
\rule[-0.5mm]{0mm}{6.5mm}
 Q  &  I  & 1048  &  725  &  -1041 &  ---  &  ---  &  2610  \\[-0.05cm] 
    &     & (997) & (678) & (-1063)&       &       & (2490) \\[0.10cm]  
 Q  &  II &  906  &  460  &  -617  &  -44  &  -143 &  2440  \\[-0.05cm] 
    &     & (866) & (429) & (-634) & (-86) & (-137)& (2316) \\[0.10cm]  
 Q  & III &  566  &  270  &  -517  &  -23  &  -82  &  2092  \\[-0.05cm] 
    &     & (545) & (249) & (-537) & (-42) & (-79) & (2013) \\[0.10cm]  
 L  &  IV &  677  &  547  &  -615  &  -32  &  -101 &  2354  \\[-0.05cm] 
    &     & (651) & (524) & (-636) & (-57) & (-98) & (2261) \\[0.10cm]  
 R  &  V  &  574  &  580  &  -610  &  -27  &  -82  &  2313  \\[-0.05cm] 
    &     & (554) & (562) & (-634) & (-45) & (-81) & (2234) \\[0.10cm]  
 E  &  VI &  396  &  722  &  -639  &  -21  &  -49  &  2288  \\[-0.05cm] 
    &     & (402) & (719) & (-682) & (-28) & (-54) & (2235)             
\end{tabular}
  \caption{The kinetic (without rest masses) and potential energy  
          contributions to the mass $M_{d'}$ of the $d'$  
          for the six parameter sets of Table I.
          The numbers in parentheses are obtained in a calculation 
          without quark exchange between the diquark and the tetraquark 
          clusters.} 
  \label{table:contributions} 
\end{center}
\end{table}

\begin{table}[htb]
\begin{center}
\begin{tabular}{ l c | c  c  c  c | c  c | c  c }
\rule[-0.5mm]{0mm}{6.5mm}Conf.\  
  & Set & $r_{D}$ & $r_{T}$ & $R_{d'}^{RGM}$ & $r_{d'} $ &
  $R_{d'}^{HO}$  & $r_{qq}^{RGM}$ & $r_{qq}^{TISM}$
\\[0.10cm]
  & & [fm] & [fm] & [fm] & [fm] & [fm] & [fm] & [fm] \\[0.15cm] \hline
\rule[-0.5mm]{0mm}{6.5mm} 
Q & I   & 0.61 & 0.74 &  1.01   & 1.16 & 0.96  & 1.32 & 1.38 \\[0.05cm]
  &     &      &      & (1.11)  &      &       &      &      \\[0.05cm]
Q & II  & 0.65 & 0.80 & 1.10    & 1.24 & 1.03  & 1.42 & 1.47 \\[0.05cm]
  &     &      &      & (1.21)  &      &       &      &      \\[0.05cm]
Q & III & 0.82 & 1.01 & 1.39    & 1.53 & 1.30  & 1.80 & 1.82 \\[0.05cm]
  &     &      &      & (1.50)  &      &       &      &      \\[0.05cm]
L & IV  & 0.74 & 0.91 & 1.32    & 1.41 & 1.18  & 1.67 & 1.73 \\[0.05cm]
  &     &      &      & (1.43)  &      &       &      &      \\[0.05cm]
R & V   & 0.81 & 0.99 & 1.46    & 1.53 & 1.27  & 1.84 & 1.86 \\[0.05cm]
  &     &      &      & (1.56)  &      &       &      &      \\[0.05cm]
E & VI  & 0.93 & 1.15 & 2.08    & 1.85 & 1.48  & 2.52 & 2.25  \\[0.05cm]
  &     &      &      & (1.94)  &      &       &      &      \\[0.05cm]
\end{tabular}
  \caption{The rms radii of the diquark ($r_{D}$), tetraquark ($r_{T}$), and
   the  intercluster wave function ($R_{d'}^{RGM}$)
   with and without (in parantheses) the quark exchange diagrams.  
   The total $d'$ matter radius $r_{d'}$ (see Eq.(21)) 
   is also given. In the RGM formalism the same value for the oscillator 
   parameter ($b_6$) is used for the diquark, tetraquark, and relative 
   motion wave functions. The mean distance between the colored clusters,
   $R_{d'}^{RGM}$, is compared to the corresponding 
   shell model radius $R_{d'}^{HO}=(15/8)\, b_6^2$. 
   The mean distance between two quarks in the RGM 
   (${r}_{qq}^{RGM}$)  and TISM (${r}_{qq}^{TISM}$) 
   approach are also shown for comparison.   
   The same notation as in Table I is used.  }
   \label{table:gcmrad}
\end{center}
\end{table}

\newcommand{\titjn}{\vec\tau_i\cdot\vec\tau_j}
\newcommand{\sisjn}{\vec\sigma_i\cdot\vec\sigma_j}
\newcommand{\liljn}{\lambda_i^a\cdot\lambda_j^a}              
\begin{table}[htb]
\begin{center}
\begin{tabular}{ l l | c c c }
\rule[-0.5mm]{0mm}{6.5mm}Diagram & Operator ${\cal{O}}_{ij}$ & $\liljn$ & 
$\liljn\sisjn$ & $\sisjn\titjn$ \\[0.15cm]
  \hline
\rule[-0.5mm]{0mm}{6.5mm}& ${\cal{O}}_{12}$ & -4/$\sqrt{3}$ &  
    8/(3$\sqrt{3}$) & 
   5/$\sqrt{3}$ \\
  direct & ${\cal{O}}_{56}$ & -8/$\sqrt{3}$ &  +8$\sqrt{3}$  & 9$\sqrt{3}$ \\
      & ${\cal{O}}_{46}$ & -2/$\sqrt{3}$ &  0 & 0 \\[0.15cm]
  \hline
      \rule[-0.5mm]{0mm}{6.5mm}& 
      ${\cal{O}}_{12}\, \hat {\rm P}_{46}^{STC}$ & 
      1/(2$\sqrt{3}$) & -2/(3$\sqrt{3}$) & -5/(4$\sqrt{3}$) \\
      & ${\cal{O}}_{34}\, \hat {\rm P}_{46}^{STC}$ & 
      -1/(2$\sqrt{3}$) & 2/(3$\sqrt{3}$) & 5/(4$\sqrt{3}$) \\
      & ${\cal{O}}_{56}\, \hat {\rm P}_{46}^{STC}$ & 0 & 0 & 0 \\          
  1q-exchange  & ${\cal{O}}_{36}\, 
                          \hat {\rm P}_{46}^{STC}$ & 
                    -1/(2$\sqrt{3}$) & 2/(3$\sqrt{3}$) & 5/(4$\sqrt{3}$) \\
      & ${\cal{O}}_{45}\, \hat {\rm P}_{46}^{STC}$ & 0 & 0 & 0 \\          
      & ${\cal{O}}_{46}\, \hat {\rm P}_{46}^{STC}$ & 
      $\sqrt{3}$/2 & 3$\sqrt{3}$/2 & 0 \\
      & ${\cal{O}}_{35}\, \hat {\rm P}_{46}^{STC}$ & 
      0 & 4/(3$\sqrt{3}$) & 5/(4$\sqrt{3}$) \\[0.15cm]
  \hline
      \rule[-0.5mm]{0mm}{6.5mm}& 
      ${\cal{O}}_{12}\, \hat {\rm P}_{46}^{STC} \hat {\rm P}_{35}^{STC}$ 
      & 1/$\sqrt{3}$ & 1/$\sqrt{3}$ & -3$\sqrt{3}$/4 \\
      & ${\cal{O}}_{34}\, \hat {\rm P}_{46}^{STC} \hat {\rm P}_{35}^{STC}$ 
      & -2/$\sqrt{3}$ & 2$\sqrt{3}$ & 9$\sqrt{3}$/4 \\
      & ${\cal{O}}_{56}\, \hat {\rm P}_{46}^{STC} \hat {\rm P}_{35}^{STC}$ 
      & -2/$\sqrt{3}$ & 2$\sqrt{3}$ & 9$\sqrt{3}$/4 \\
   2q-exchange  & ${\cal{O}}_{23}\, 
                          \hat {\rm P}_{46}^{STC} \hat {\rm P}_{35}^{STC}$ 
      & -5/(4$\sqrt{3}$) & 0 & 0 \\
      & ${\cal{O}}_{25}\, \hat {\rm P}_{46}^{STC} \hat {\rm P}_{35}^{STC}$ 
      & -5/(4$\sqrt{3}$) & 0 & 0 \\
      & ${\cal{O}}_{46}\, \hat {\rm P}_{46}^{STC} \hat {\rm P}_{35}^{STC}$ 
      & 1/(4$\sqrt{3}$) & 0 & -$\sqrt{3}$/4 \\
      & ${\cal{O}}_{45}\, \hat {\rm P}_{46}^{STC} \hat {\rm P}_{35}^{STC}$ 
      & 1/(4$\sqrt{3}$) & 0 & -$\sqrt{3}$/4 
\end{tabular}
  \caption{Spin(S)-Isospin(T)-Color(C) matrix elements of the 
interaction kernels}  
  \label{table:stcelemente}
\end{center}
\end{table}

\vfill
\eject

\newpage



\vspace{0.5cm}
\begin{figure}[ht]
\caption{The formation of a compound six-quark system from the 
initial $\pi^+ nn$ system. 
The $\bar d$ quark in the $\pi^+$ annihilates with a $d$ quark in one of the 
neutrons leading to a $q^2$--$q^4$ clusterized six-quark state.
In the stringlike bag model, colored diquark and tetraquark clusters are bound
together by a color-electric flux-tube (string), 
which is rotating with
angular momentum $L=1$. In the present model, the relative $q^2-q^4$ wave 
function is calculated from the Hamiltonian of Eq.(1).} 
\label{fig1}
\begin{center}
  {\epsfig{file=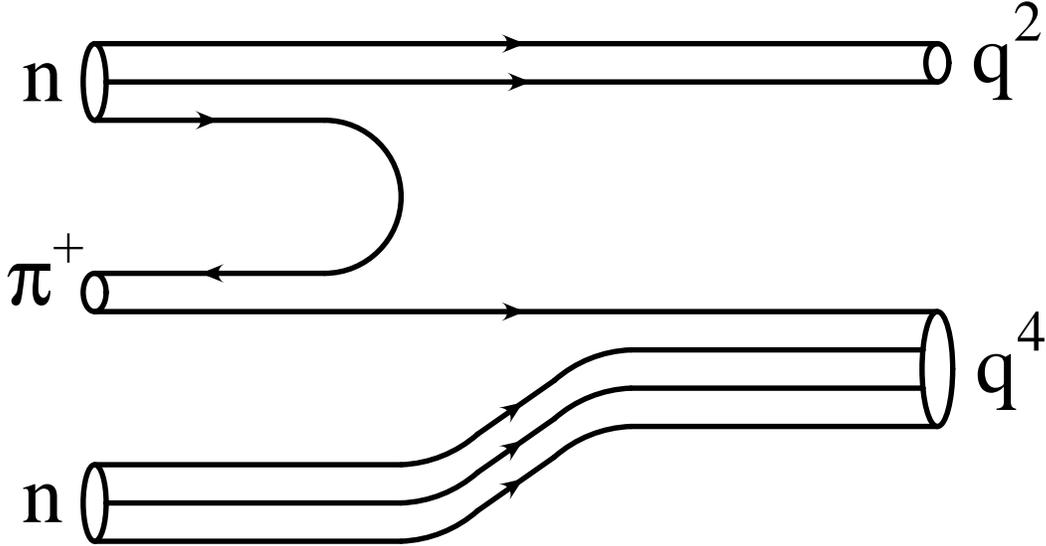, width=14.5cm}}
\end{center}
\end{figure}


\vspace{0.5cm}
\begin{figure}[ht]
\caption{The
 residual (a) one-gluon, (b) one-pion,  and (c) one-sigma exchange 
interactions. The finite size of the quark-meson vertex described by the 
cut-off $\Lambda$ is indicated by a small black dot.}
\begin{center}
  {\epsfig{file=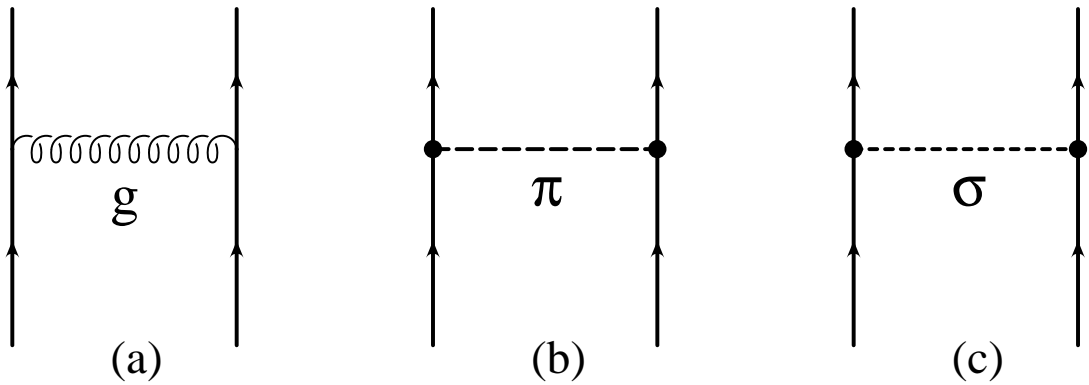, width=14.5cm}}
\end{center}
\label{fig2}
\end{figure}

\newpage 


\vspace{0.5cm}
\begin{figure}[ht]
\caption{The radial dependence of typical confinement models in 
the nucleon.} 
\begin{center}
  {\epsfig{file=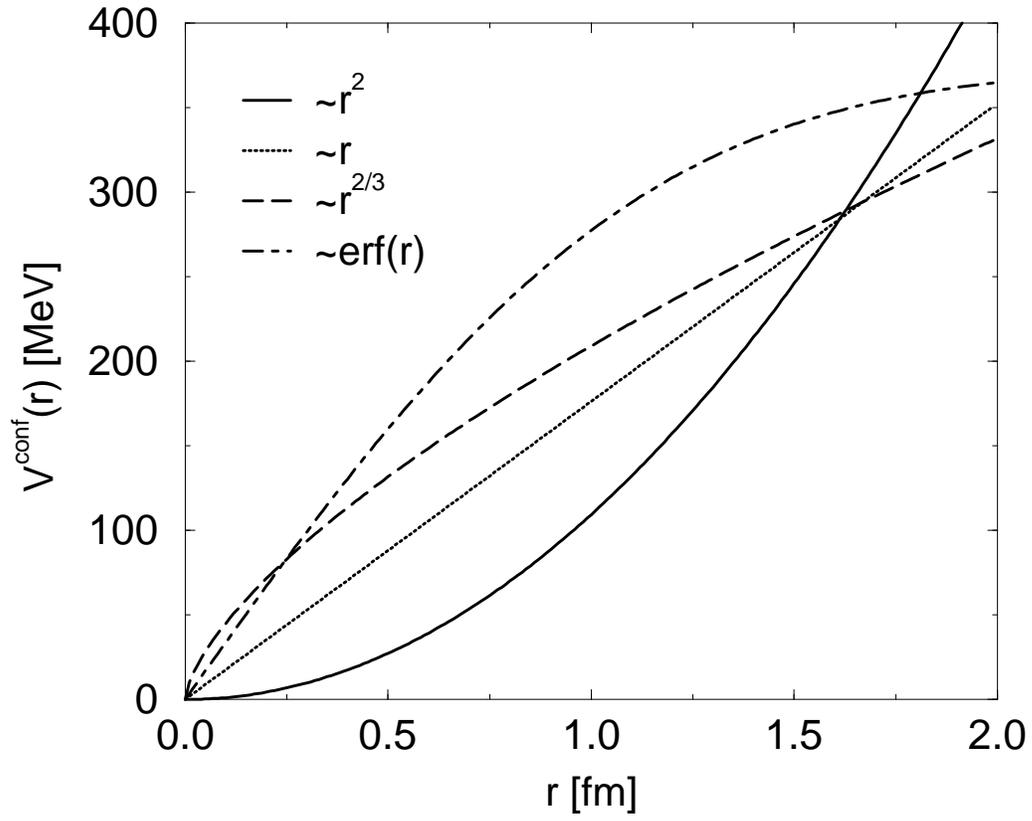, width=14.5cm}}
\end{center}
\label{fig3}
\end{figure}

\newpage 


\vspace{0.5cm}
\begin{figure}[ht]
\caption{The Jacobi coordinates in the $q^2$--$q^4$  colored cluster model 
of the $d'$.} 
\begin{center}
  {\epsfig{file=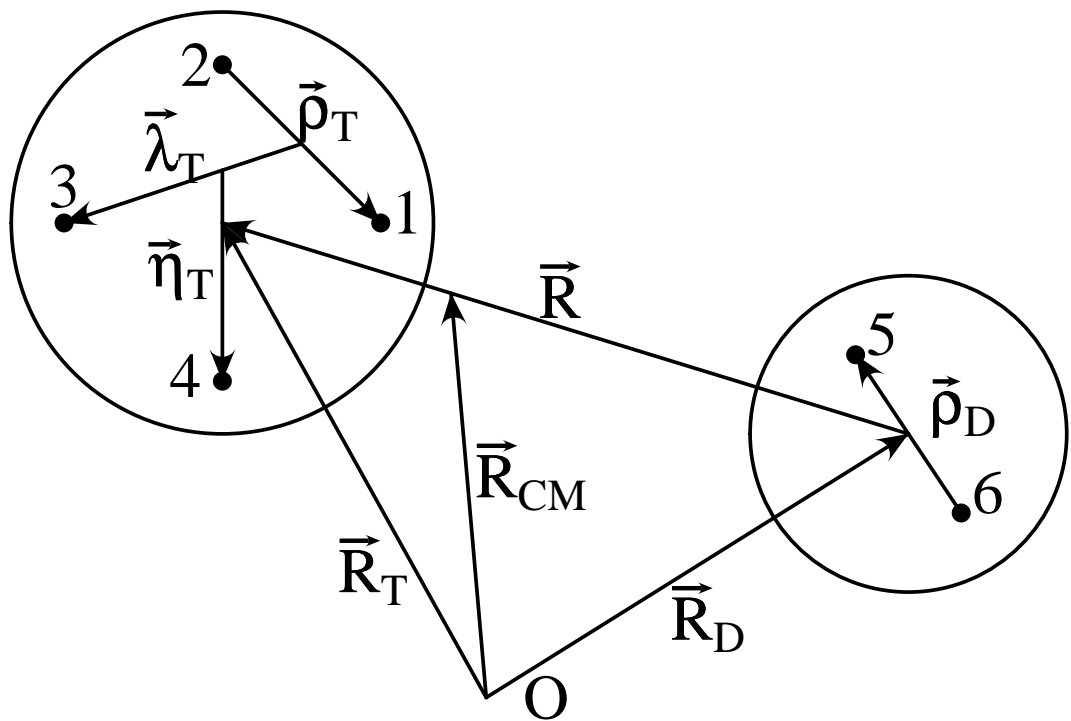, width=14.5cm}}
\end{center}
\label{fig4}
\end{figure}
 
\newpage 


\vspace{0.5cm}
\begin{figure}[ht]
\caption{The
 direct, one-quark $V_{ij}\,P_{46}^{OSTC}$ 
and two-quark  $V_{ij}\,P_{46}^{OSTC}\, P_{35}^{OSTC}$
exchange diagrams for the one-gluon 
exchange potential.
Corresponding diagrams are calculated for the confinement, one-pion, 
and one-sigma exchange potentials. 
The stringlike bag model includes only the 
diagrams in the first row. The quark exchange diagrams neccessary to 
satisfy the Pauli principle for the whole six-quark system are neglected
in the stringlike bag model but included in the present theory.
The quark exchange kernels increase the $d'$ mass by about 
100 MeV.} 
\begin{center}
  {\epsfig{file=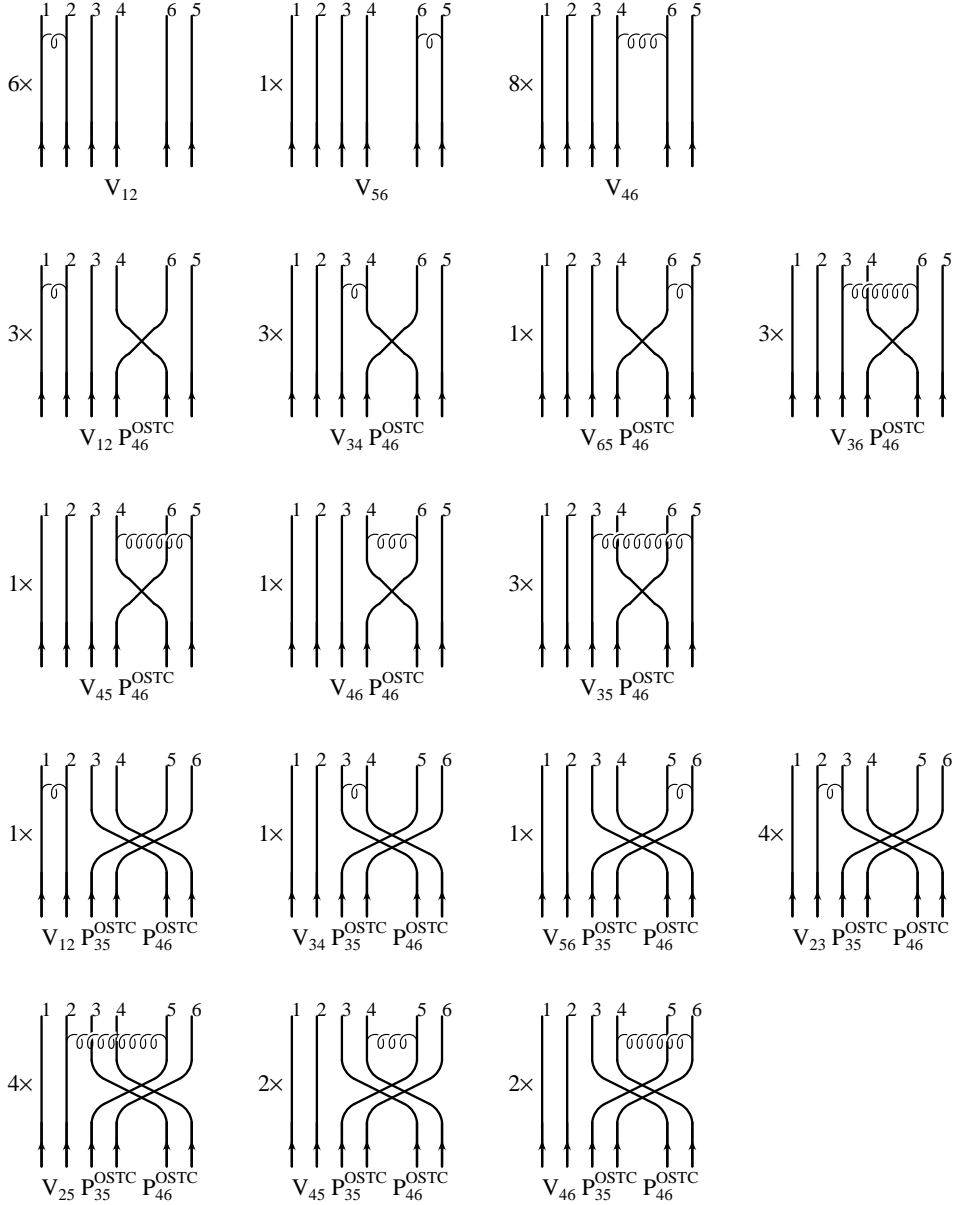, width=15.5cm}}
\end{center}
\label{fig5}
\end{figure}

\newpage 


\vspace{0.5cm}
\begin{figure}[ht]
\caption{The relative wave RGM 
function between the tetraquark and diquark clusters 
with (Pauli-on) and without
(Pauli-off) inclusion of the quark exchange diagrams for the 
smallest (model II)  and largest (model VI) $d'$-dibaryon. The RGM wave 
functions are compared to the pure harmonic oscillator wave function of 
Eq.\ (23). The upper graph shows the results for the 
quadratic confinement (model II), while the lower graph  is obtained for 
the color-screened 
confinement (model VI). For model II (upper graph) the clusters  
overlap, while they are just touching in model VI. The corresponding 
rms radii are given in Table IV.
The wave functions of the other models are in between
these two extremes. } 
\begin{center}
  {\epsfig{file=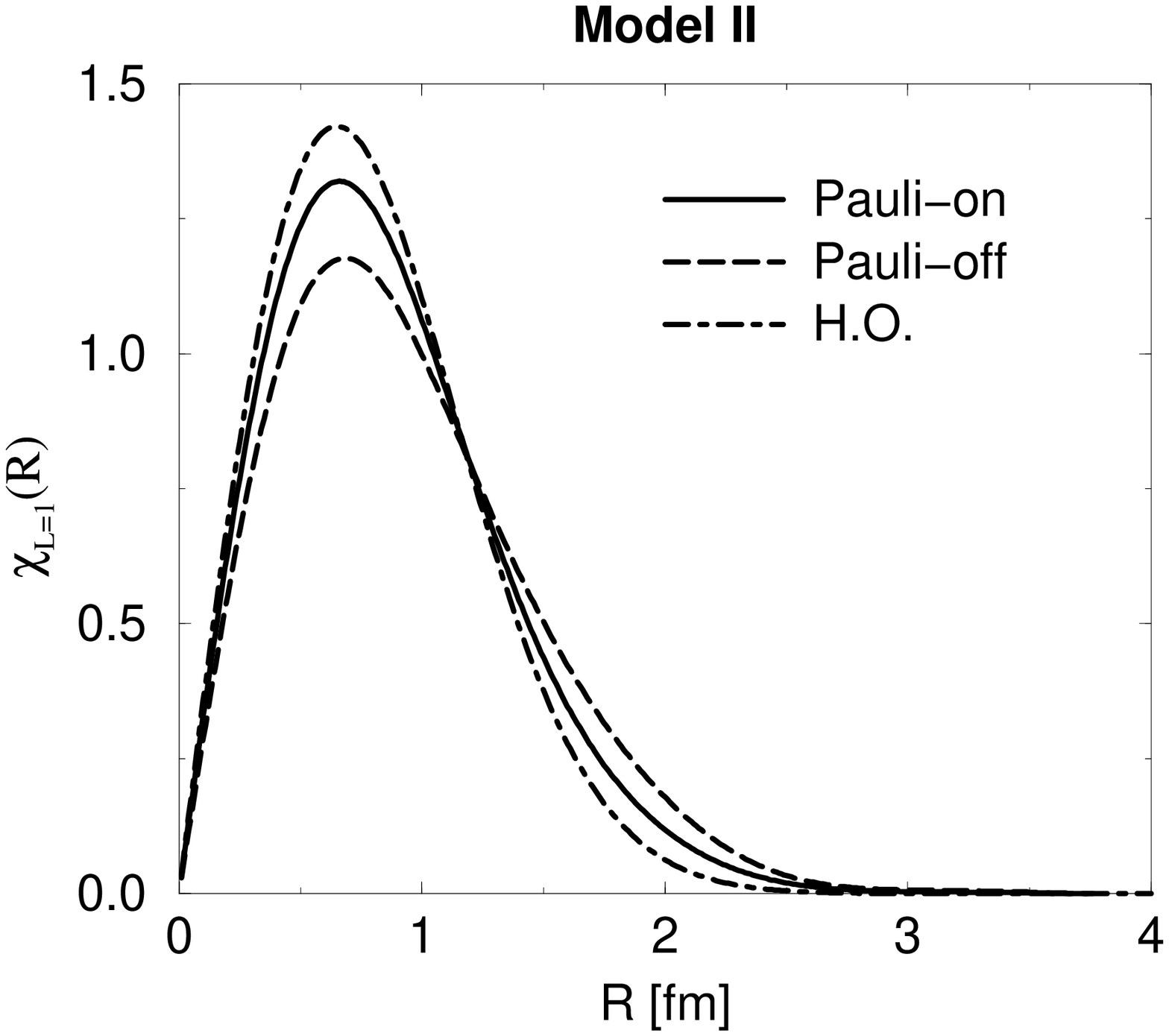, width=11.5cm,height=8.0cm}}
  {\epsfig{file=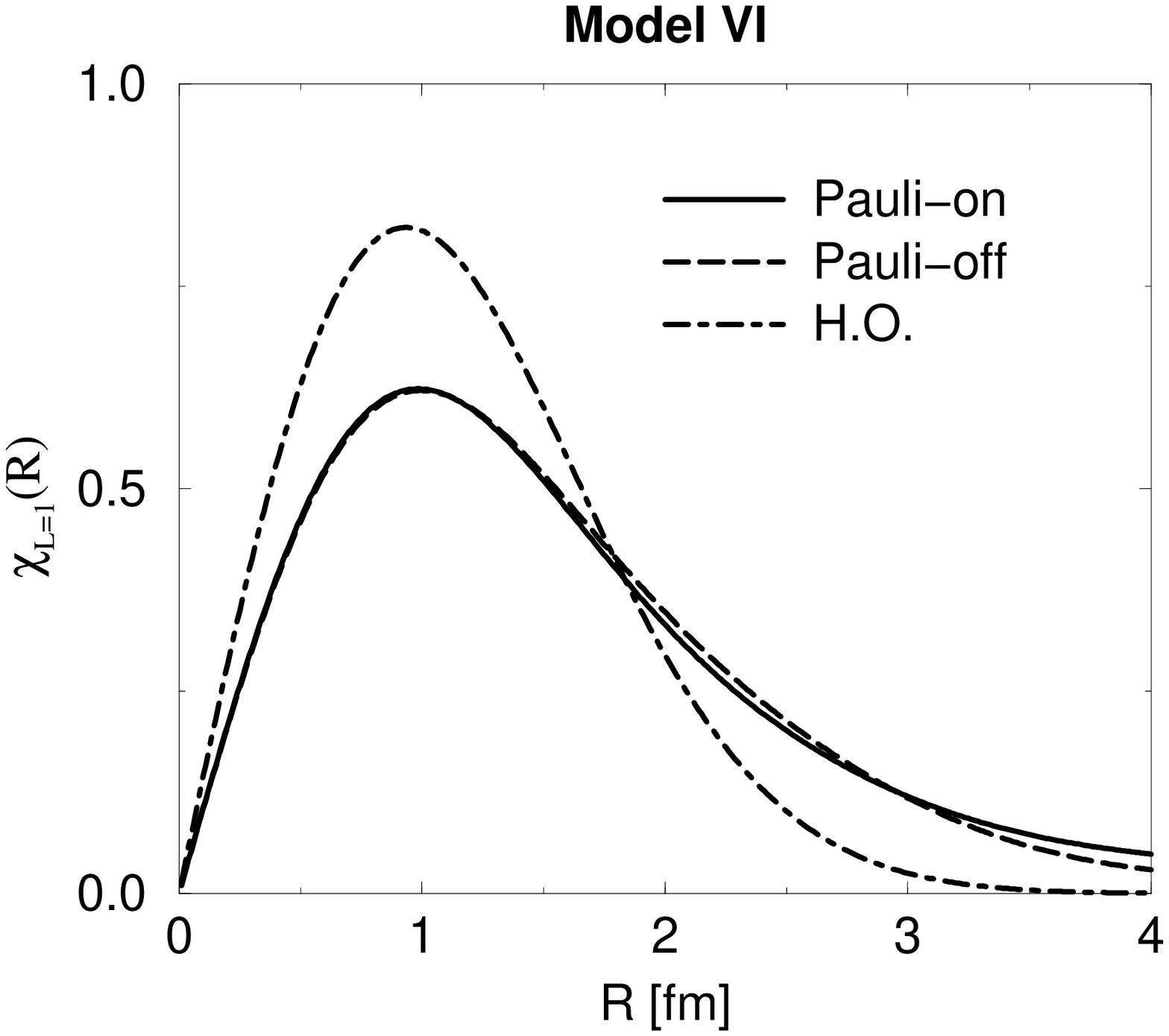, width=11.5cm,height=8.0cm}}
\end{center}
\label{fig6}
\end{figure}


\begin{thebibliography}{99}
\bibitem{com1} 
It has been argued that because quark exchange between the clusters 
corresponds to a quark-antiquark exchange process in the crossed channel, 
there could be some overlap with the direct meson exchange interaction between 
quarks belonging to different clusters. However, it has previously been shown 
that for scalar ($\sigma$) and pseudoscalar ($\pi$) meson exchanges between 
quarks, the quark exchange diagrams can be added to 
the direct intercluster meson exchange diagrams without danger of double 
counting \cite{Yaz90}. Also in a field-theoretical description 
of $NN$ scattering in the quark model using Low's equations 
one finds that the two-quark exchange diagrams can 
be clearly distinguished from the colorless $q{\bar q}$ (meson) exchange 
diagrams \cite{Mac97}.
\bibitem{com2} 
We do not attempt to justify this rather large factor of three here.
This is clearly an artefact of the harmonic oscillator confinement model
and should be taken with some caution in other applications.
We expect that $a_c^{(3)}/a_c^{(6)}$ will be substantially smaller for
more realistic confinement models.
\end{thebibliography}
\end{document}